\documentclass[
  aps,
  prd,
  reprint,
  superscriptaddress,
  nofootinbib,
  longbibliography
]{revtex4-2}

\usepackage{fontspec}
\usepackage{amsmath,amssymb}
\usepackage{bm}
\usepackage{placeins}
\usepackage{graphicx}
\usepackage{textcase}
\usepackage{booktabs}
\usepackage{multirow}

\newcommand{\fithead}[2]{%
  \begin{tabular}[c]{@{}c@{}}#1\\[-1pt]#2\end{tabular}%
}
\usepackage{array}
\usepackage{tabularx}
\usepackage{makecell}
\usepackage{siunitx}

\usepackage{adjustbox}
\usepackage{rotating}
\usepackage{changepage}
\usepackage{xcolor}
\usepackage{slashed}
\usepackage{tikz-feynman}
\usetikzlibrary{calc,positioning,shapes.geometric}
\tikzfeynmanset{compat=1.1.0,warn luatex=false}
\allowdisplaybreaks
\renewcommand{\arraystretch}{1.15}

\tikzset{
  baryon/.style={line width=0.8pt},
  decuplet/.style={double, double distance=1.8pt, line width=0.8pt},
  meson/.style={dash pattern=on 4pt off 4pt, line width=0.8pt},
  weak/.style={
    draw,
    rectangle,
    fill=white,
    line width=1.2pt,
    minimum size=7.5pt,
    inner sep=0pt
  },
  strong/.style={circle, fill=black, minimum size=7.5pt, inner sep=0pt}
}

\usepackage[colorlinks=true,linkcolor=blue,citecolor=blue,urlcolor=blue]{hyperref}

\begin{document}

\title{\texorpdfstring{
$|\Delta I|=3/2$ non-leptonic hyperon decays in covariant baryon chiral perturbation theory
}{Delta I = 3/2 nonleptonic hyperon decays}}

\author{Jie Xu}
\affiliation{School of Physics, Beihang University, Beijing 100191, China}
\author{Jun-Xu Lu}
\affiliation{School of Physics, Beihang University, Beijing 100191, China}

\author{Rui-Xiang Shi}
\email[Corresponding author: ]{ruixiang.shi@gxnu.edu.cn}
\affiliation{Department of Physics, Guangxi Normal University, Guilin 541004, China}
\affiliation{Guangxi Key Laboratory of Nuclear Physics and Technology, Guangxi Normal University, Guilin 541004, China}

\author{Li-Sheng Geng}
\email[Corresponding author: ]{lisheng.geng@buaa.edu.cn}
\affiliation{School of Physics,  Beihang University, Beijing 102206, China}
\affiliation{Sino-French Carbon Neutrality Research Center, \'Ecole Centrale de P\'ekin/School of General Engineering, Beihang University, Beijing 100191, China}
\affiliation{Peng Huanwu Collaborative Center for Research and Education, Beihang University, Beijing 100191, China}
\affiliation{Southern Center for Nuclear-Science Theory (SCNT), Institute of Modern Physics, Chinese Academy of Sciences, Huizhou 516000, Guangdong Province, China}

\date{March 2026}

\begin{abstract}
Inspired by the recent BESIII measurements of non-leptonic hyperon decays, we reexamine their $|\Delta I|=3/2$ amplitudes in
covariant baryon chiral perturbation theory with the extended-on-mass-shell renormalization scheme. Using the same restricted set of diagrammatic topologies as in the early analyses in heavy baryon
chiral perturbation theory, we assess the effects of relativistic corrections, explicit decuplet baryons, different spin-$3/2$ coupling schemes, and pion-loop contributions. Our results show that relativistic effects alone lead to only mild changes, while decuplet contributions, especially in the consistent-coupling scheme, significantly improve the fit quality. Pion-loop contributions further reduce the $\chi^2$ values. We highlight the importance of the consistent coupling scheme in the decuplet sector for describing the selected $|\Delta I|=3/2$ amplitudes.

\end{abstract}

\maketitle

\section{Introduction}

Non-leptonic hyperon decays provide a vital experimental and theoretical venue for exploring the nature of
non-perturbative strong interactions, testing the elusive $|\Delta I|=1/2$ rule in strangeness-changing processes, and searching for potential new sources of CP violation~\cite{GellMannPais1955Glasgow,Dalitz1963DeltaI12,Suzuki1965DeltaI12Hyperon,HeStegerValencia1991,Chang:2000hu,TandeanValencia2003,Holstein:1999gx}. For the hyperons of the lowest-lying spin-1/2 baryon octet, there are seven distinct decay channels, i.e., $\Lambda\to p\pi^-$, $\Lambda\to n\pi^0$, $\Sigma^+\to p\pi^0$, $\Sigma^+\to n\pi^+$, $\Sigma^-\to n\pi^-$, $\Xi^0\to\Lambda\pi^0$, and $\Xi^-\to\Lambda\pi^-$. These decays receive contributions from both $|\Delta I|=1/2$ and $|\Delta I|=3/2$ transitions, with their amplitudes fully described by $S$- and $P$-wave components, which correspond to odd and even parity, respectively.

As is well established experimentally~\cite{Bangerter1966SigmaDecay,Thorne1989DeltaIHalf}, the $|\Delta I|=1/2$ amplitudes dominate the nonleptonic decays of hyperons. A longstanding puzzle associated with these decays concerns the $S$- and $P$-wave components, with theories typically describing either the $S$- or $P$-wave data but not both simultaneously~\cite{Holstein:1999gx}. Considerable theoretical efforts, employing phenomenological models~\cite{LeYaouanc1979Hyperon,Donoghue1981BagModel,ManoharGeorgi1984ChiralQuarkModel,Donoghue1986LowEnergyWeak,DonoghueHolstein1986,FloresMendieta2019SWave} and chiral perturbation theory~($\chi$PT)~\cite{Bijnens1985,Jenkins1992,BorasoyHolstein1999,BorasoyHolstein1999Resonance,AbdElHadyTandean2000HyperonNLChPTReexamined,ZhangYang2025,Salone2026}, have been devoted to this issue. In contrast, the $|\Delta I|=3/2$ sector remains largely unexplored theoretically, primarily owing to the paucity and low precision of the available experimental data, as well as the large theoretical uncertainties inherent in the calculations. In Refs.~\cite{HeValencia1997,AbdElHady1999DeltaI32}, the authors investigated the $|\Delta I|=3/2$ non-leptonic decays of hyperons in heavy baryon chiral perturbation theory~(HB $\chi$PT) at tree and one-loop levels, respectively. However, the experimental data fitted in those works are now obsolete, and the contributions of certain Feynman diagrams were neglected.

In recent years, the BESIII Collaboration has developed an innovative method that exploits quantum correlations and cascade decay properties of baryon-antibaryon pairs to investigate non-leptonic decays of $\Lambda$, $\Sigma$ and $\Xi$ hyperons~\cite{BESIII:2021ypr,BESIII:2018cnd,BESIII:2020fqg,BESIII:2022qax,BESIII:2023drj,BESIII:2023sgt}. The measured asymmetry parameter $\alpha$ associated with the $|\Delta I|=3/2$ amplitudes for the $\Xi$ hyperon deviates from the previous PDG average by about $2\sigma$. More remarkably, for the $\Lambda$ hyperon, the deviation reaches a significance level exceeding $5\sigma$, consistent with the CLAS experiment~\cite{Ireland:2019uja}. Such a substantial shift in the asymmetry parameters necessitates re-extracting the experimental data for the $|\Delta I|=3/2$ amplitudes and calls for an improved theoretical description of non-leptonic hyperon decays in baryon chiral perturbation theory.

As a low-energy effective field theory of QCD, baryon chiral perturbation theory can be formulated in three distinct variants, depending on how to treat power-counting-breaking terms: the HB $\chi$PT in the non‑relativistic framework~\cite{JenkinsManohar1991,Jenkins1992Mass,BernardKaiserMeissner1995ChiralDynamics}, the covariant baryon chiral perturbation theory with the infrared regularization~(IR)~\cite{BecherLeutwyler1999,KubisMeissner2001BaryonFormFactorsIR,KubisMeissner2001NucleonFormFactorsIR} and the extended on‑mass‑shell renormalization~(EOMS)~\cite{Fuchs2003Renormalization,Fuchs2003VectorMesons} schemes, respectively. Among these, the EOMS scheme has proven successful in tackling a series of long‑standing problems in low‑energy nonperturbative strong‑interaction physics, including baryon magnetic moments~\cite{Geng2008MagneticMoments,Xiao2018OctetMagneticMomentsNNLO,Shi2019CharmedMagneticMoments}, proton Compton scattering~\cite{Lensky:2009uv}, pion-proton scattering~\cite{Alarcon:2012kn,Chen:2012nx}, nucleon-nucleon interaction~\cite{Lu:2021gsb}, and resonant antikaon-nucleon scattering~\cite{Lu:2022hwm}. Furthermore, it exhibits faster convergence than the HB and IR formulations~\cite{Geng2013CovariantSU3}. Notably, no covariant effective‑field‑theory study has yet been devoted to the $|\Delta I|=3/2$ non-leptonic hyperon decays. Therefore, applying the EOMS $\chi$PT to these processes is highly desirable.

In the present work, we utilize the EOMS $\chi$PT to study the $|\Delta I|=3/2$ non-leptonic hyperon decays with constraints from the recent BESIII measurements. It should be emphasized that the purpose of this work is not to provide a complete phenomenological description of the $|\Delta I|=3/2$ amplitudes, because a full next-to-leading order~(NLO) treatment would require many additional low-energy constants~(LECs). Rather, we aim to evaluate the impact of relativistic corrections, different treatments of the decuplet- and octet-baryon interactions with the
pseudoscalar mesons, and the pion‑loop contributions.  

This work is organized as follows. In Sec.~\ref{Sec2}, we provide the relevant Lagrangians, power counting rule, and computations of the $|\Delta I|=3/2$ amplitudes. Results and discussions are
given in Sec.~\ref{Sec3}, followed by a summary
in Sec.~\ref{Sec4}.

\section{Theoretical framework}\label{Sec2}
\subsection{Lagrangians}

We first introduce the chiral Lagrangians relevant to this study, which are
\begin{equation}
    \mathcal{L}
    =
    \mathcal{L}_{\mathrm{str}}
    +
    \mathcal{L}_{\mathrm{w}},
\end{equation}
where $\mathcal{L}_{\mathrm{str}}$ denotes the strong chiral Lagrangian and $\mathcal{L}_{\mathrm{w}}$ represents the weak chiral Lagrangian constructed using the spurion method.
Expanding these Lagrangians according to the chiral power counting~\cite{Fuchs2003Renormalization,Scherer2003ChPT}, we write
\begin{align}
    \mathcal{L}_{\mathrm{str}}
    &=
    \mathcal{L}^{(1)}_{\phi B}
    +
    \mathcal{L}^{(1)}_{\phi D}
    +
    \mathcal{L}^{(1)}_{\phi BD}
    +
    \cdots,
    \\
    \mathcal{L}_{\mathrm{w}}
    &=
    \mathcal{L}^{W(1)}_{\phi B}
    +
    \mathcal{L}^{W(1)}_{\phi D}
    +
    \cdots.
\end{align}
The numbers in the parentheses denote the corresponding orders in the chiral power counting. 
The ellipses represent higher-order terms not needed in the present work.

For the strong interaction involving only the octet baryons,  the leading-order meson-baryon Lagrangian is given by~\cite{Jenkins1992Mass,Jenkins1992,AbdElHady1999DeltaI32,Scherer2003ChPT,Salone2026}:
\begin{equation}
\begin{aligned}
    \mathcal{L}_{\phi \mathrm{B}}^{(1)}
=&\left\langle\bar{B}\left(i \slashed{D}-m_{0}\right) B\right\rangle
+\frac{D }{2}\left\langle\bar{B} \gamma^{\mu} \gamma_{5}\{u_{\mu}, B\}\right\rangle\\
&+\frac{F }{2}\left\langle\bar{B} \gamma^{\mu} \gamma_{5}[u_{\mu}, B]\right\rangle,
\label{eq:chiral_OctetLagrangian}
\end{aligned}
\end{equation}
where $\langle\cdots\rangle$ denotes the trace in flavor space, $D_\mu$ is the chiral covariant derivative, $m_0$ denotes the octet-baryon mass in the chiral limit, and $u^{\mu}=i\left(u \partial^{\mu} u^{\dagger}-u^{\dagger} \partial^{\mu} u\right)$. 
The pseudoscalar Goldstone fields are collected in the $SU(3)$ matrix
\begin{equation}\label{eq:chiral_u}
    U(\phi)=u^2(\phi)=\exp \left\{\frac{i \phi}{F_{0}}\right\},
\end{equation}
 with 
\begin{equation}\label{eq:chiral_phi}
    \phi=\sqrt{2}\left(\begin{array}{ccc}
\frac{1}{\sqrt{2}} \pi^{0}+\frac{1}{\sqrt{6}} \eta & \pi^{+} & K^{+} \\
\pi^{-} & -\frac{1}{\sqrt{2}} \pi^{0}+\frac{1}{\sqrt{6}} \eta & K^{0} \\
K^{-} & \bar{K}^{0} & -\frac{2}{\sqrt{6}} \eta
\end{array}\right),
\end{equation}
where $F_{0}$ denotes the pseudoscalar decay constant in the chiral limit.
The
octet baryons are collected in the matrix $B$,
\begin{equation}\label{eq:chiral_B}
    B=\left(\begin{array}{ccc}
\frac{1}{\sqrt{2}} \Sigma^{0}+\frac{1}{\sqrt{6}} \Lambda & \Sigma^{+} & p \\
\Sigma^{-} & -\frac{1}{\sqrt{2}} \Sigma^{0}+\frac{1}{\sqrt{6}} \Lambda & n \\
\Xi^{-} & \Xi^{0} & -\frac{2}{\sqrt{6}} \Lambda
\end{array}\right).
\end{equation}
The covariant derivative appearing in the octet kinetic term is defined as
\begin{align}
  & \mathcal{D}^{\mu}B
=
\partial^{\mu}B
+
\left[\Gamma^{\mu},B\right],\\
&\Gamma^{\mu}
=
\frac{1}{2}
\left(
u\partial^{\mu}u^{\dagger}
+
u^{\dagger}\partial^{\mu}u
\right).
\end{align}
For the strong interaction involving the decuplet baryons, we first consider the conventional-coupling Lagrangian~\cite{AbdElHady1999DeltaI32,Jenkins1992,Pascalutsa1998,PascalutsaTimmermans1999,Salone2026},
\begin{align}
\mathcal{L}_{\phi B D}^{ (1)}
&=\frac{\mathcal{C}}{2F_{0}} \varepsilon^{a b c} \bar{T}_{\mu}^{a d e} B_{c}^{e} \partial^{\mu} \phi_{b}^{d}
+\text{H.c.},
\label{eq:chiral_BDMLagrangian_conventional}
\\
\mathcal{L}_{\phi D D}^{(1)}
&=\frac{\mathcal{H}}{2F_{0}}
\bar{T}_{a b c}^{\mu} \gamma_{\nu} \gamma_{5}\partial^{\nu} \phi_{d}^{c} T_{\mu}^{a b d}.
\label{eq:chiral_DDMLagrangian_conventional}
\end{align}
The decuplet baryons are described by the Rarita--Schwinger field
$T_{abc}^{\mu}$, which satisfies the constraint
$\gamma_{\mu}T_{abc}^{\mu}=0$ and is completely symmetric in its
$SU(3)$ flavor indices $a$, $b$, and $c$. Its components are
\begin{equation}
\begin{array}{lll}
T_{111}=\Delta^{++},
&
T_{112}=\frac{1}{\sqrt{3}}\Delta^{+},
&
T_{122}=\frac{1}{\sqrt{3}}\Delta^{0},
\\[2mm]
T_{222}=\Delta^{-},
&
T_{113}=\frac{1}{\sqrt{3}}\Sigma^{*+},
&
T_{123}=\frac{1}{\sqrt{6}}\Sigma^{*0},
\\[2mm]
T_{223}=\frac{1}{\sqrt{3}}\Sigma^{*-},
&
T_{133}=\frac{1}{\sqrt{3}}\Xi^{*0},
&
T_{233}=\frac{1}{\sqrt{3}}\Xi^{*-},
\\[2mm]
T_{333}=\Omega^{-},
&
&
\end{array}
\end{equation}
where the Lorentz index has been suppressed.

The conventional-coupling scheme, which has been widely used in studies of non-leptonic hyperon decays~\cite{Jenkins1992,AbdElHady1999DeltaI32,Salone2026}, inevitably introduces unphysical spin‑1/2 contributions.
To avoid these unphysical components, one may instead adopt the consistent-coupling scheme, in which the effective Lagrangians are invariant under spin-$3/2$ gauge transformations~\cite {Pascalutsa1998,PascalutsaTimmermans1999}. 
In this formulation, the  spin-$1/2$ components can be absorbed into suitable higher-order low-energy constants. 
The corresponding Lagrangians read~\cite{Pascalutsa1998,PascalutsaTimmermans1999,Geng2009Decuplet,Ren2014Decuplet}:
\begin{align}
\mathcal{L}_{\phi B D}^{\prime(1)}
&=\frac{i C}{2m_{D} F_{0}} \varepsilon^{a b c}
\left(\partial_{\alpha} \bar{T}_{\mu}^{a d e}\right)
\gamma^{\alpha \mu \nu} B_{c}^{e} \partial_{\nu} \phi_{b}^{d}
+\text{H.c.},
\label{eq:chiral_BDMLagrangian_consistent}
\\
\mathcal{L}_{\phi D D}^{\prime(1)}
&=\frac{-i \mathcal{H}}{2m_{D} F_{0}}
\bar{T}_{\mu}^{a b c} \gamma^{\mu \nu \rho \sigma} \gamma_{5}
\left(\partial_{\rho} T_{\nu}^{a b d}\right) \partial_{\sigma} \phi_{d}^{c}.
\label{eq:chiral_DDMLagrangian_consistent}
\end{align}
Here, $m_D$ is the chiral limit mass of the decuplet baryons in the chiral limit. The overall minus sign in Eq.~\eqref{eq:chiral_DDMLagrangian_consistent} is introduced to match the sign convention in Eq.~\eqref{eq:chiral_DDMLagrangian_conventional}.
The totally antisymmetric products of gamma matrices are defined by
\begin{align}
\gamma^{\mu\nu}
&=
\frac{1}{2}
[\gamma^\mu,\gamma^\nu],
\\
\gamma^{\mu\nu\rho}
&=
\frac{1}{2}
\{\gamma^{\mu\nu},\gamma^\rho\},
\\
\gamma^{\mu\nu\rho\sigma}
&=
\frac{1}{2}
[\gamma^{\mu\nu\rho},\gamma^\sigma].
\end{align}

We now turn to the weak sector. 
At the quark level, the $|\Delta S|=1$, $|\Delta I|=3/2$ transitions are described by the four-quark effective weak Hamiltonian
\begin{equation}
\label{eq:weak_quark_Hamiltonian}
\mathcal{H}_{\mathrm{eff}}^{(27_L,1_R)}
=
\frac{G_F}{\sqrt{2}}
V_{ud}^* V_{us}
\left(
\frac{c_1+c_2}{3}
\right)
\mathcal{O}_{|\Delta I|=3/2}^{(27_L,1_R)}
+
\text{H.c.}.
\end{equation}
The operator $\mathcal{O}_{|\Delta I|=3/2}^{(27_L,1_R)}$ transforms as $(27_L,1_R)$ under $SU(3)_L\times SU(3)_R$ and is realized in the chiral theory through the spurion tensor $T_{ij,kl}$.
The corresponding leading-order weak chiral Lagrangians~\cite{HeValencia1997,AbdElHady1999DeltaI32} read:
\begin{align}
\mathcal{L}^{W(1)}_{\phi B}
&=
\beta_{27}
T_{ij,kl}
\left(
u \bar{B} u^\dagger
\right)_{ki}
\left(
u B u^\dagger
\right)_{lj}
+
\text{H.c.},
\label{eq:weak_3/2_lagrangian_BB}
\\
\mathcal{L}^{W(1)}_{\phi D}
&=
\delta_{27}
T_{ij,kl}
u_{kd}
u^\dagger_{bi}
u_{le}
u^\dagger_{cj}
\left(
\bar{T}^\mu
\right)_{abc}
\left(
T_\mu
\right)_{ade}
+
\text{H.c.}.
\label{eq:weak_3/2_lagrangian_DD}
\end{align}
The nonzero elements of $T_{ij,kl}$ that project out the 
$|\Delta S|=1$, $|\Delta I|=3/2$ component are~\cite{AbdElHady1999DeltaI32,Georgi1984}
\begin{equation}
\label{eq:T_matrix_elements}
\begin{aligned}
T_{12,13}
&=
T_{21,13}
=
T_{12,31}
=
T_{21,31}
=
\frac{1}{2},
\\
T_{22,23}
&=
T_{22,32}
=
-\frac{1}{2}.
\end{aligned}
\end{equation}

\subsection{Power counting breaking and restoration}
In $\chi$PT, the relative importance of different Feynman diagrams is determined by their chiral orders, $\nu$, with the corresponding contributions scaling as $(p/\Lambda_{\chi})^{\nu}$~\cite{Weinberg1979,Scherer2003ChPT}.
For processes in the one-baryon sector, i.e., with a single baryon in both the initial and final states, the chiral order of a diagram is given by
\begin{equation}
\nu = 4L + \sum_n n V_n - 2N_M - N_B \, .
\end{equation}
Here, $L$ denotes the number of loops, $V_n$ is the number of vertices of order $n$, and $N_M$ and $N_B$ represent the numbers of internal meson and baryon lines, respectively.

The chiral expansion is organized in powers of small quantities, which are assigned the following chiral orders:
\begin{equation}
\begin{aligned}
m_{\pi},\, m_{K},\, m_{\eta} &\sim \mathcal{O}(p), \\
m_{D}-m_{0}=\delta &\sim \mathcal{O}(p), \\
m_{B} - m_{0} &\sim \mathcal{O}(p^2) ,
\end{aligned}
\end{equation}
where $m_{B}$ denotes the physical mass of the external baryon, and the mass splitting $\delta$ is treated following the small-scale expansion~\cite{Hemmert1998ChiralDelta}.\footnote{In contrast, Jenkins and Hady count $\delta$ as of $\mathcal{O}(p^2)$~\cite{Jenkins1992,AbdElHady1999DeltaI32}. We have verified that this difference in the chiral counting of $\delta$ does not affect our conclusions.}

It is worth noting that, for the baryon propagators appearing in the $P$-wave pole diagrams shown in Fig.~\ref{fig:tree_diagrams}$(b1)$ and $(b2)$, as well as the baryon propagator outside the loop in Fig.~\ref{fig:loop_pwave}, the standard power counting~\cite{Fuchs2003Renormalization,Scherer2003ChPT} treats the denominator $\slashed{p}-m_0$ as of $\mathcal{O}(p)$. 
However, when the external baryons are put on-shell, the denominator reduces to $m_{B} - m_{0}$, which formally scales as $\mathcal{O}(p^2)$. 
In the present work, we follow the standard power counting and count these denominators as of $\mathcal{O}(p)$.

In the pure mesonic sector, the chiral expansion is organized in powers of external momenta and Goldstone-boson masses, and loop diagrams obey the standard power-counting rules. 
In the baryonic sector, the baryon mass remains finite in the chiral limit, and loop amplitudes may therefore contain lower-order analytic pieces that violate the naive power counting. 
These terms are referred to as power-counting-breaking (PCB) terms~\cite{Fuchs2003Renormalization,Du2016PCB}.
The extended-on-mass-shell (EOMS) renormalization scheme restores a consistent chiral power counting by subtracting the PCB terms from the loop amplitudes;
equivalently, these terms are absorbed into the corresponding LECs~\cite{GegeliaJaparidze1999,GegeliaJaparidzeWang1999,Fuchs2003Renormalization,Fuchs2003VectorMesons,Scherer2003ChPT,Geng2013CovariantSU3}.
In the present work, the usual $\widetilde{\mathrm{MS}}$ subtraction is performed together with the finite EOMS subtraction.

\subsection{\texorpdfstring{$|\Delta I|=3/2$ amplitudes}{|ΔI|=3/2 amplitudes}}
The amplitudes for non-leptonic hyperon decays have the following form~\cite{Bijnens1985,Jenkins1992,AbdElHady1999DeltaI32}:
\begin{equation}\label{eq:IM}
    iM_{{B_{i}\rightarrow B_{f}\pi}}=G_{F}m_{\pi^{+}}^{2}\bar{u}_{B_{f}}(A^{S}_{B_{f}B_{i}}+A^{P}_{B_{f}B_{i}}\gamma_{5})u_{B_{i}},
\end{equation}
where $A^{S}_{B_{f}B_{i}}$ and $A^{P}_{B_{f}B_{i}\pi}$ denote the $S$-wave and $P$-wave amplitudes,  respectively. Isospin symmetry implies the following relations 
\begin{subequations}\label{eq:3/2_isospin_relations}
\renewcommand{\theequation}{\theparentequation\textit{\alph{equation}}}
\begin{align}
\mathcal{M}_{\Sigma^{+} \rightarrow n \pi^{+}}
-\sqrt{2} \mathcal{M}_{\Sigma^{+} \rightarrow p \pi^{0}}
+2 \mathcal{M}_{\Sigma^{-} \rightarrow n \pi^{-}} & =0, \\
\mathcal{M}_{\Lambda \rightarrow n \pi^{0}}
-\sqrt{2} \mathcal{M}_{\Lambda \rightarrow p \pi^{-}} & =0, \\
\mathcal{M}_{\Xi^{0} \rightarrow \Lambda \pi^{0}}
-\sqrt{2} \mathcal{M}_{\Xi^{-} \rightarrow \Lambda \pi^{-}} & =0.
\end{align}    
\end{subequations}
These isospin relations given in Eq.~\eqref{eq:3/2_isospin_relations} hold separately for the $S$- and $P$-wave decay amplitudes, so that only four independent amplitudes need to be evaluated. Following Refs.~\cite{Jenkins1992,AbdElHady1999DeltaI32,BorasoyHolstein1999}, we choose the four channels of $\Sigma^{+}\rightarrow n\pi^{+}$, $\Sigma^{-}\rightarrow n\pi^{-}$, $\Lambda\rightarrow p\pi^{-}$, and $\Xi^{-}\rightarrow \Lambda \pi^{-}$.

The general structure of the amplitudes for $|\Delta I|=3/2$ non-leptonic hyperon decays can be written as
\begin{subequations}\label{eq:iM_structure}
\renewcommand{\theequation}{\theparentequation\textit{\alph{equation}}}
\begin{align}
A^{S}_{B_{f}B_{i}}
&=
\frac{1}{\sqrt{2}F_{\pi}}
\left\{
\alpha^{S}_{B_{f}B_{i}}
\left(1+\lambda_{B_{f}B_{i}\pi}\right)
+\bar{\Sigma}^{S}_{B_{f}B_{i}}
\right\},
\label{iM_S_structure}
\\
A^{P}_{B_{f}B_{i}}
&=
\frac{1}{\sqrt{2}F_{\pi}}
\left\{
\alpha^{P}_{B_{f}B_{i}}
\left(1+\lambda_{B_{f}B_{i}\pi}\right)
+\bar{\Sigma}^{P}_{B_{f}B_{i}}
+\gamma^{P}_{B_{f}B_{i}}
\right\}.
\label{iM_P_structure}
\end{align}
\end{subequations}
The amplitudes in Eq.~\eqref{eq:iM_structure} consist of tree-level and one-loop corrections. The coefficients $\alpha^{S,P}_{B_f B_i}$ denote the tree-level contributions. The factor $\lambda_{B_f B_i\pi}$ collects the corrections from wave-function renormalization and from the replacement of the pseudoscalar decay constant in the chiral limit, $F_{0}$, by the physical pion decay constant, $F_\pi$. The factor $\lambda_{B_f B_i \pi}$ is given by
\begin{equation}
\lambda_{B_{f}B_{i}\pi}
=\frac{1}{2}\left(Z_{B_{f}}-1\right)
+\frac{1}{2}\left(Z_{B_{i}}-1\right)
+\frac{1}{2}\left(Z_{\pi}-1\right)
+\delta F_{\pi},
\end{equation}
where the explicit expressions for the wave-function renormalization factors $Z$ and the correction $\delta F_{\pi}$ are given in Appendix~\ref{app:wfr}.

\begin{figure}[!t]
\centering

\includegraphics[width=0.5\textwidth]{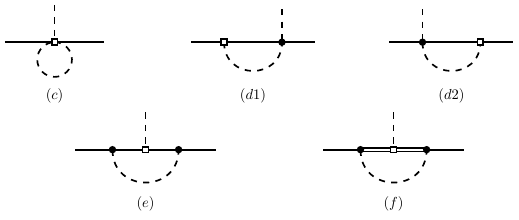}

\caption{One-loop diagrams contributing to $S$-wave hyperon non-leptonic decay amplitudes. Double (single) solid lines denote baryon decuplet (octet) fields.}
\label{fig:loop_swave}
\end{figure}

\begin{figure}[!t]
\centering

\includegraphics[width=0.45\textwidth]{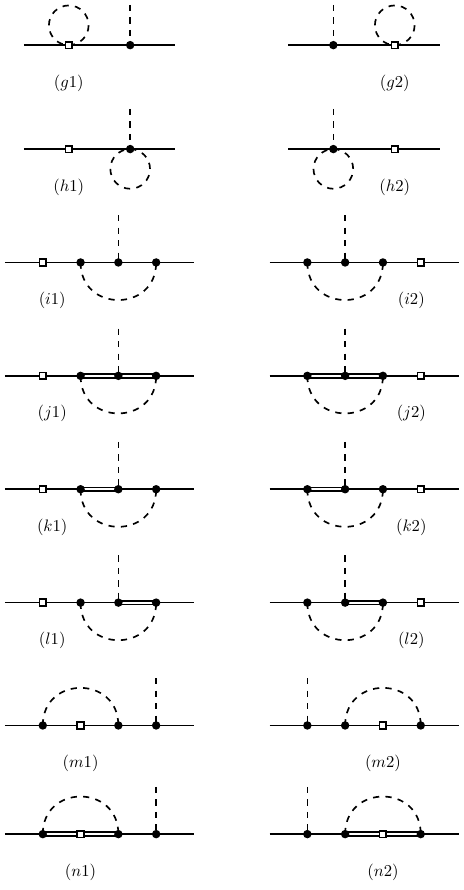}

\caption{One-loop diagrams contributing to $P$-wave hyperon non-leptonic decay amplitudes.}
\label{fig:loop_pwave}
\end{figure}

The quantities $\bar{\Sigma}^{S,P}_{B_f B_i}$ represent the contributions from the one-loop diagrams and can be decomposed as
\begin{equation}
\bar{\Sigma}^{S,P}_{B_f B_i}
=
\Sigma^{S,P}_{B_f B_i}
+
\Sigma^{\prime\,S,P}_{B_f B_i},
\end{equation}
where $\Sigma^{S,P}_{B_f B_i}$ arises from loop diagrams with only octet baryons in the intermediate states, while $\Sigma^{\prime\,S,P}_{B_f B_i}$ corresponds to loop diagrams involving decuplet baryons. The corresponding Feynman diagrams are shown in Figs.~\ref{fig:loop_swave} and~\ref{fig:loop_pwave}.

The term $\gamma^{P}_{B_f B_i}$, which appears only in the $P$-wave amplitudes, originates from loop corrections to the baryon propagators in the pole diagrams and can be similarly decomposed into octet and decuplet contributions. The corresponding Feynman diagrams are shown in Fig.~\ref{fig:loop_extra_pwave}. 

\begin{figure}[!t]
\centering

\includegraphics[width=0.5\textwidth]{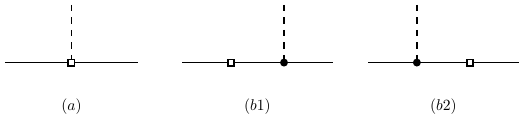}

\caption{Tree-level diagrams for $(a)$ $S$-wave and $(b1,b2)$ $P$-wave hyperon non-leptonic decays. Solid (dashed) lines denote baryon-octet (meson-octet) fields. Solid dots (hollow squares) represent strong (weak) vertices, derived from the lagrangians  $\mathcal{L}_{\mathrm{str}}$ and $\mathcal{L}_{\mathrm{w}}$.}
\label{fig:tree_diagrams}
\end{figure}

\begin{figure}[!t]
\centering

\includegraphics[width=0.45\textwidth]{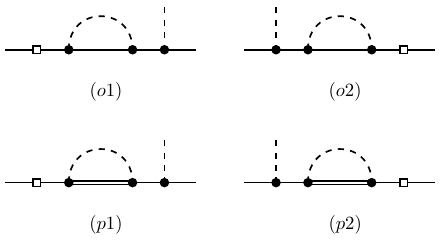}

\caption{Feynman diagrams for the loop corrections to the baryon propagators in the $P$-wave pole diagrams shown in Fig.~\ref{fig:tree_diagrams}(b1) and (b2).} 
\label{fig:loop_extra_pwave}
\end{figure}

For these propagators in Fig.~\ref{fig:tree_diagrams} and outside the loop in Fig.~\ref{fig:loop_pwave}, their  denominators can reduce to the following form
\[
\slashed{p}-m_0 \;\to\; m_B-m_0,
\]
with  the on-shell condition for the external baryon. 
However, the parameter $m_0$ is not well constrained in practice, and the resulting amplitudes are sensitive to its value, particularly in the $P$-wave sector. 
To reduce this sensitivity and improve the stability of the predictions, we adopt a pragmatic prescription~\cite{Springer1995Revisited,BorasoyHolstein1999}.
Specifically, for these baryon propagators, the mass parameter $m_0$ appearing in the denominator is replaced by the physical mass of the corresponding intermediate baryon.
This prescription reduces the sensitivity introduced by the poorly constrained
chiral-limit baryon masses.

This mass replacement, however, requires a separate treatment of the
propagator-correction diagrams shown in
Fig.~\ref{fig:loop_extra_pwave}. This is because the part of these diagrams
that generates the replacement of the mass parameter in the pole-diagram
propagator with the corresponding physical baryon mass has already been
incorporated into the tree-level results in
Eq.~\eqref{eq:tree_level_p_wave_amplitudes}. To avoid double counting, only the remaining contributions need to be included explicitly. The details of this treatment are given in
Appendix~\ref{Treatment of propagator corrections in the P-wave pole diagrams}.

It should be noted that the loop diagrams in Fig.~\ref{fig:loop_swave}-\ref{fig:loop_extra_pwave}  are not exhaustive of all NLO topologies. To facilitate comparison with earlier HB $\chi$PT analyses~\cite{AbdElHady1999DeltaI32}, we restrict our calculation to the same diagrams as used therein, which is one of the purposes of this work. We stress that our work does not provide a complete phenomenological description of non-leptonic hyperon decays, but rather an assessment of the impact of relativistic effects, different decuplet coupling schemes, and pion-loop contributions. We have checked that the remaining one-loop diagrams shown in Appendix~\ref{app:unconsidered_p2_loops}  vanish in the heavy-baryon limit.

The Feynman diagrams contributing to $S$- and $P$-wave amplitudes at  $\mathcal{O}(1)$ are shown in Fig.~\ref{fig:tree_diagrams}. These leading-order contributions of $\mathcal{O}(1)$ are calculated by the Lagrangians of Eq.~(\ref{eq:weak_3/2_lagrangian_BB}) and Eq.~(\ref{eq:chiral_OctetLagrangian}), and the corresponding results are given as follows:
\vspace{-0.5em}
\begin{equation}
\begin{aligned}
\alpha^{S}_{\Sigma^{+}n}
=\frac{3}{2}\beta_{27},
\quad
\alpha^{S}_{\Sigma^{-}n}
=-\beta_{27},
\quad
\alpha^{S}_{\Lambda n}
=0,
\quad
\alpha^{S}_{\Xi^{-}\Lambda}
=0.
\end{aligned}
\label{eq:tree_level_s_wave_amplitudes}
\end{equation}
\vspace{-1em}
\begin{subequations}\label{eq:tree_level_p_wave_amplitudes}
\renewcommand{\theequation}{\theparentequation\textit{\alph{equation}}}
\begin{align}
\alpha^{P}_{\Sigma^{+}n}
&=
-\frac{\beta_{27}(D+3F)(m_{\Sigma}+m_{N})}
{2(m_{\Sigma}-m_{N})},
\label{eq:tree_level_p_sigma_plus}
\\
\alpha^{P}_{\Sigma^{-}n}
&=
\frac{\beta_{27}F(m_{\Sigma}+m_{N})}
{m_{\Sigma}-m_{N}},
\label{eq:tree_level_p_sigma_minus}
\\
\alpha^{P}_{\Lambda n}
&=
\frac{\beta_{27}D(m_{\Lambda}+m_{N})}
{\sqrt{6}(m_{\Sigma}-m_{N})},
\label{eq:tree_level_p_lambda}
\\
\alpha^{P}_{\Xi^{-}\Lambda}
&=
-\frac{\beta_{27}D(m_{\Xi}+m_{\Lambda})}
{\sqrt{6}(m_{\Xi}-m_{\Sigma})}.
\label{eq:tree_level_p_xi}
\end{align}
\end{subequations}
where these results agree with those obtained in HB $\chi$PT~\cite{AbdElHady1999DeltaI32}. \footnote{When comparing with Ref.~\cite{AbdElHady1999DeltaI32}, one should account for an overall minus sign arising from sign conventions.}

We now turn to the NLO contributions arising from one-loop diagrams shown in Figs.~\ref{fig:loop_swave} and~\ref{fig:loop_pwave}. The explicit loop-integral expressions are given in Appendix~\ref{app:amplitudes}. At this order, the decay amplitudes receive NLO analytic contributions from both one-loop diagrams and local counterterms. The LECs appearing in the local counterterms, however, cannot be determined from the limited experimental data. The earlier HB $\chi$PT studies~\cite{Bijnens1985,Jenkins1992,AbdElHady1999DeltaI32,AbdElHadyTandean2000HyperonNLChPTReexamined} therefore focused on the NLO non-analytic contributions and omitted the local counterterms. In the same spirit, the present work subtracts the NLO analytic pieces generated by the covariant loop amplitudes and neglects the local counterterms. The remaining loop contributions contain the NLO nonanalytic terms, together with formally higher-order relativistic corrections. We then assess the impact of different theoretical ingredients within the
EOMS scheme by comparing the quality of the corresponding fits. We also
perform fits in which the loop-generated NLO analytic terms are retained to
assess the sensitivity of our conclusions to these terms.

The loop amplitudes are evaluated with the aid of \texttt{FeynCalc}, yielding expressions in terms of the Passarino--Veltman scalar functions $A_{0}$, $B_{0}$, and $C_{0}$~\cite{PassarinoVeltman1979OneLoopCorrections}. To keep only the NLO non-analytic pieces, we perform a chiral expansion for the EOMS loop amplitudes and isolate the NLO analytic terms to subtract.
Since the three-point function $C_{0}$ does not possess an analytic expression under general kinematic conditions, the chiral expansion of the loop amplitudes cannot be performed explicitly. We therefore adopt a simplifying approximation in which the baryon masses are set to a common value of $1~\mathrm{GeV}$, and the outgoing pion mass is set to zero, allowing us to derive analytic expressions for the NLO subtraction terms. Note that this approximation is used solely for this purpose.


\section{Results and discussions}\label{Sec3}
For each physical decay channel, we use the convention
\begin{equation}
s = A^{S}_{B_{f}B_{i}}, \qquad
p = \frac{|\vec{k}|}{E_{B_{f}}+M_{B_{f}}}A^{P}_{B_{f}B_{i}},
\end{equation}
where $\vec{k}$ is the three-momentum of the outgoing pion in the rest frame of the initial baryon.

The theoretical $|\Delta I|=3/2$ isospin $S$-wave amplitudes are obtained from the corresponding particle-basis amplitudes as
\vspace{-0.5em}
\begin{equation}\label{eq:3/2_isospin_relation}
\begin{aligned}
\quad\quad S_{3,3}^{(\Lambda)}&=\frac{\sqrt{2}}{3}\left(\sqrt{2}\, s_{\Lambda \rightarrow n \pi^{0}}+s_{\Lambda \rightarrow p \pi^{-}}\right), \\
\quad\quad 
 S_{3,1}^{(\Xi)}&=\frac{\sqrt{2}}{3}\left(\sqrt{2}\, s_{\Xi^{0} \rightarrow \Lambda \pi^{0}}+s_{\Xi^{-} \rightarrow \Lambda \pi^{-}}\right), \\
S_{3,3}^{(\Sigma)}&=-\frac{1}{3}\left(s_{\Sigma^{+} \rightarrow n \pi^{+}}-\sqrt{2}\, s_{\Sigma^{+} \rightarrow p \pi^{0}}-s_{\Sigma^{-} \rightarrow n \pi^{-}}\right).
\end{aligned}
\end{equation}
The corresponding $P$-wave amplitudes are obtained by replacing $S$ and $s$ with $P$ and $p$ in Eq.~\eqref{eq:3/2_isospin_relation}, respectively. Since final-state interactions are not included in the present calculation, no phase-shift factors are introduced.
The data of $|\Delta I|=3/2$ isospin amplitudes are taken from Ref.~\cite{Salone2026} and listed in Table~\ref{tab:deltaI32_input}.
\begin{table}[!htbp]
\caption{Data of $|\Delta I|=3/2$ isospin amplitudes, taken from Ref.~\cite{Salone2026}.
}
\label{tab:deltaI32_input}
\begin{ruledtabular}
{\renewcommand{\arraystretch}{1.35}
\begin{tabular}{l S[table-format=-1.3(2), separate-uncertainty=false]}
Observable & {Value} \\
\colrule
$S_{3,3}^{(\Lambda)}$ & -0.045(9)  \\
$S_{3,1}^{(\Xi)}$     & -0.072(18) \\
$S_{3,3}^{(\Sigma)}$  & -0.035(5)  \\
$P_{3,3}^{(\Lambda)}$ &  0.71(13)  \\
$P_{3,1}^{(\Xi)}$     &  0.27(13)  \\
$P_{3,3}^{(\Sigma)}$  & -0.83(6)   \\
\end{tabular}}
\end{ruledtabular}
\end{table}

In the previous HB $\chi$PT study~\cite{AbdElHady1999DeltaI32}, the NLO non-analytic contributions of amplitudes $S_{3}^{(\Lambda)}$ and $S_{3}^{(\Xi)}$ is zero. 
For a direct comparison between the EOMS and HB results about the NLO non-analytic terms, we exclude $S_{3}^{(\Lambda)}$ and $S_{3}^{(\Xi)}$ and restrict our fits to the remaining four observables:
\begin{equation}
\left\{
S_{3}^{(\Sigma)},\;
P_{3}^{(\Lambda)},\;
P_{3}^{(\Xi)},\;
P_{3}^{(\Sigma)}
\right\}.
\end{equation}

\begin{table*}[!htbp]
\centering
\footnotesize
\setlength{\tabcolsep}{1.6pt}
\renewcommand{\arraystretch}{1.20}

\caption{
Fit results for the $|\Delta I|=3/2$ $S$- and $P$-wave isospin amplitudes in the setup where the loop-generated analytic pieces are subtracted.  
Here ``O" denotes the inclusion of octet-baryon loop contributions in addition to the tree-level contributions, while ``D" denotes the further inclusion of decuplet-baryon loop contributions, evaluated using the conventional (conv.) and consistent (consist.) coupling schemes. 
The columns labeled ``O+D+$\pi$" additionally include pion-loop contributions. 
The uncertainties assigned to the input values used in the fits are shown in the ``Expt." column and include the uncertainties extracted from fits to experimental data, quoted in Ref.~\cite{Salone2026}, as well as the estimated truncation uncertainty and the uncertainty associated with NLO analytic contributions not included in the fits. 
For the observables predicted by the present fits, the uncertainties are estimated by propagating the fitted-LEC uncertainties while neglecting correlations among the fitted LECs. The LECs are given in units of $\sqrt{2}F_{\pi}G_{F}m_{\pi^{+}}^{2}$.
}
\label{tab:updated_hyperon_amplitudes}

\makebox[\textwidth][c]{%
\begin{tabular}{
@{}
l
c
>{\hspace{2em}}c<{\hspace{-0.5em}}
c
ccccc
c
@{}
}
\toprule

& \multirow{2}{*}{Tree level}
& \multicolumn{2}{c}{HB with NLO Analytic Terms omitted}
& \multicolumn{5}{c}{EOMS with NLO Analytic Terms omitted}
& \multirow{2}{*}{Expt.} \\

\cmidrule(lr){3-4} \cmidrule(lr){5-9}

&
& O
& \fithead{O+D}{(conv.)}
& O
& \fithead{O+D}{(conv.)}
& \fithead{O+D+$\pi$}{(conv.)}
& \fithead{O+D}{(consist.)}
& \fithead{O+D+$\pi$}{(consist.)}
& \\

\midrule

\addlinespace[1pt]
\multicolumn{10}{c}{\textbf{Observables}} \\
\addlinespace[1pt]

$S^{\Sigma}_{3,3}$ 
& $\phantom{-}0.070(106)$ 
& $\phantom{-}0.100(94)$ 
& $\phantom{-}0.017(248)$ 
& $\phantom{-}0.109(93)$ 
& $-0.019(130)$ 
& $-0.052(119)$ 
& $-0.133(146)$ 
& $-0.170(109)$
& $-0.035(115)$ \\

$P^{\Lambda}_{3,3}$ 
& $-0.014(21)$ 
& $-0.027(25)$ 
& $\phantom{-}0.001(51)$ 
& $-0.033(28)$ 
& $\phantom{-}0.004(20)$ 
& $\phantom{-}0.002(5)$ 
& $-0.002(13)$ 
& $-0.059(18)$
& $\phantom{-}0.710(220)$ \\

$P^{\Xi}_{3,1}$ 
& $\phantom{-}0.038(58)$ 
& $\phantom{-}0.095(89)$ 
& $\phantom{-}0.264(339)$ 
& $\phantom{-}0.104(89)$ 
& $\phantom{-}0.277(227)$ 
& $\phantom{-}0.282(213)$ 
& $\phantom{-}0.340(228)$ 
& $\phantom{-}0.353(171)$
& $\phantom{-}0.270(174)$ \\

$P^{\Sigma}_{3,3}$ 
& $-0.027(41)$ 
& $-0.052(49)$ 
& $\phantom{-}0.000(92)$ 
& $-0.068(58)$ 
& $\phantom{-}0.002(27)$ 
& $-0.009(5)$ 
& $-0.118(47)$ 
& $-0.355(123)$
& $-0.830(216)$ \\

\midrule

\addlinespace[1pt]
\multicolumn{10}{c}{\textbf{LECs}} \\
\addlinespace[1pt]

$\beta_{27}$ 
& $-0.070(106)$ 
& $-0.051(48)$ 
& $\phantom{-}0.054(73)$ 
& $-0.053(45)$ 
& $\phantom{-}0.024(53)$ 
& $\phantom{-}0.064(114)$ 
& $\phantom{-}0.101(56)$ 
& $\phantom{-}0.241(93)$
& \\

$\delta_{27}$ 
& -- 
& -- 
& $\phantom{-}1.185(880)$ 
& -- 
& $\phantom{-}0.397(261)$ 
& $\phantom{-}0.407(244)$ 
& $\phantom{-}0.405(176)$ 
& $\phantom{-}0.320(143)$
& \dots \\

\midrule

$\chi^2$ 
& $27.25$ 
& $26.56$ 
& $25.36$ 
& $26.31$ 
& $25.13$ 
& $24.85$ 
& $22.23$ 
& $18.63$
& \\

\bottomrule
\end{tabular}%
}

\end{table*}

\begin{table*}[!htbp]
\centering
\footnotesize
\setlength{\tabcolsep}{3.2pt}
\renewcommand{\arraystretch}{1.20}

\caption{
Same as Table \ref{tab:updated_hyperon_amplitudes}, but with the loop-generated analytic pieces retained. 
}
\label{tab:analytic_terms_amplitudes}

\makebox[\textwidth][c]{%
\begin{tabular}{@{}l ccccc c@{}}
\toprule

& \multicolumn{5}{c}{EOMS With NLO Loop-generated Analytic Terms retained}
& \multirow{2}{*}{Expt.} \\

\cmidrule(lr){2-6}

& O
& \fithead{O+D}{(conv.)}
& \fithead{O+D+$\pi$}{(conv.)}
& \fithead{O+D}{(consist.)}
& \fithead{O+D+$\pi$}{(consist.)}
& \\

\midrule

\addlinespace[1pt]
\multicolumn{7}{c}{\textbf{Observables}} \\
\addlinespace[1pt]

$S^{\Sigma}_{3,3}$ 
& $\phantom{-}0.088(104)$ 
& $-0.045(116)$ 
& $\phantom{-}0.060(49)$ 
& $-0.182(117)$ 
& $\phantom{-}0.056(29)$
& $-0.035(115)$ \\

$P^{\Lambda}_{3,3}$ 
& $-0.020(23)$ 
& $-0.001(3)$ 
& $-0.121(91)$ 
& $\phantom{-}0.012(10)$ 
& $-0.122(40)$
& $\phantom{-}0.710(220)$ \\

$P^{\Xi}_{3,1}$ 
& $\phantom{-}0.052(61)$ 
& $\phantom{-}0.272(173)$ 
& $\phantom{-}0.289(362)$ 
& $\phantom{-}0.317(193)$ 
& $\phantom{-}0.249(254)$
& $\phantom{-}0.270(174)$ \\

$P^{\Sigma}_{3,3}$ 
& $-0.043(51)$ 
& $-0.004(8)$ 
& $-0.198(149)$ 
& $-0.164(74)$ 
& $-0.669(218)$
& $-0.830(216)$ \\

\midrule

\addlinespace[1pt]
\multicolumn{7}{c}{\textbf{LECs}} \\
\addlinespace[1pt]

$\beta_{27}$ 
& $-0.056(66)$ 
& $\phantom{-}0.044(91)$ 
& $\phantom{-}0.304(230)$ 
& $\phantom{-}0.175(87)$ 
& $\phantom{-}0.477(154)$
& \\

$\delta_{27}$ 
& -- 
& $\phantom{-}0.412(262)$ 
& $-0.026(416)$ 
& $\phantom{-}0.487(218)$ 
& $-0.175(262)$
& \dots \\

\midrule

$\chi^2$ 
& $26.97$ 
& $25.08$ 
& $23.51$ 
& $21.27$ 
& $15.49$
& \\

\bottomrule
\end{tabular}%
}

\end{table*}

\begin{figure*}[!t]
    \centering
    \includegraphics[width=0.9\textwidth]{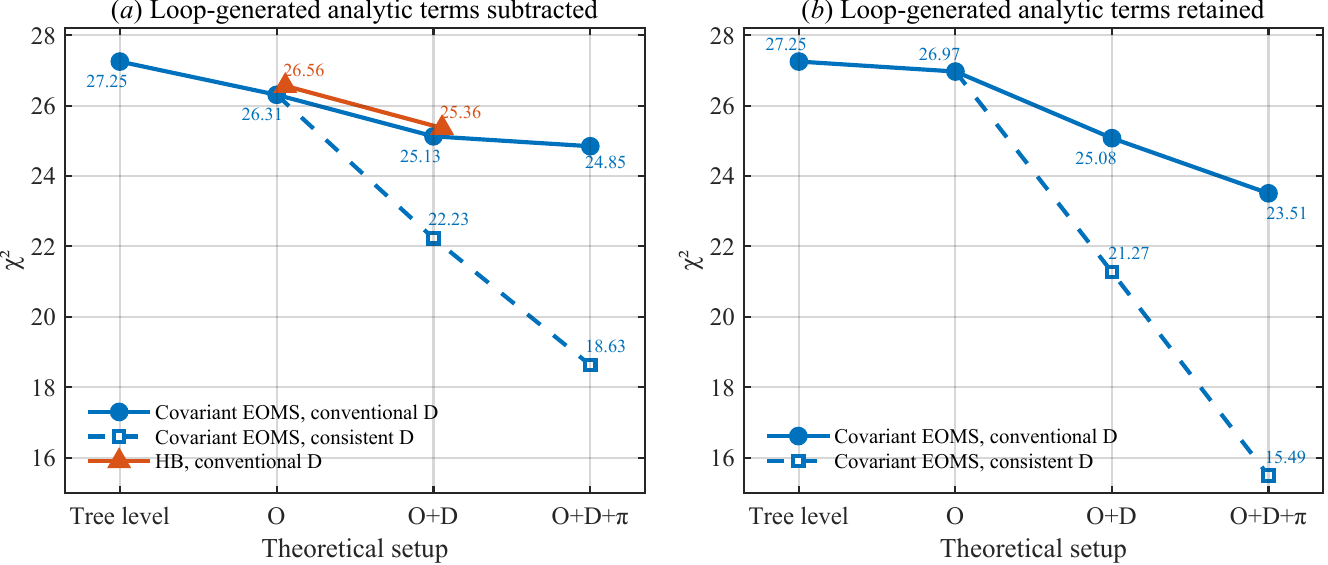}
    \caption{
Evolution of the $\chi^2$ values for the $|\Delta I|=3/2$ amplitude fits under different theoretical setups. 
Panel $(a)$ shows the results obtained after subtracting the loop-generated analytic pieces, while panel $(b)$ shows the corresponding covariant EOMS results with these analytic pieces retained. 
The horizontal axis indicates the successive inclusion of octet loops (O), decuplet intermediate states (D), and pion-loop contributions. 
For the decuplet contributions, the conventional and consistent coupling schemes are shown separately. 
The common tree-level and octet-loop stages are displayed before the conventional and consistent branches split. 
The numerical $\chi^2$ values shown in the subfigures are taken from Tables~\ref{tab:updated_hyperon_amplitudes} and~\ref{tab:analytic_terms_amplitudes}.
}
    \label{fig:observable_merged}
\end{figure*}
For the numerical analysis, we fix the strong LECs to values commonly used in the literature. 
Following Refs.~\cite{Jenkins1992,Granados:2017jnp,Salone2026}, we take
\begin{equation}
\mathcal{C}=1.6, 
\quad 
\mathcal{H}=-1.9 \pm 0.7, 
\quad 
D=0.80, 
\quad 
F=0.46 .
\end{equation}
For the octet-baryon mass in the chiral limit, we use $m_0 = 0.88~\mathrm{GeV}$,
as determined in Ref.~\cite{Ren2012}. 
The octet-decuplet mass splitting is chosen as $\delta = m_D - m_0 = 0.231~\mathrm{GeV}$~\cite{Geng2009Decuplet}.
The subtraction scale is set to $\mu = 1~\mathrm{GeV}$.

We follow the strategies of Refs.~\cite{Salone2026,Jenkins1992} for estimating the truncation error of the chiral expansion and the uncertainty associated with NLO analytic contributions not included in the fits, respectively. 
For each fitted amplitude $A_i$, the truncation uncertainty is approximated as
\begin{equation}
\sigma_{{\rm trunc},i}
=
\frac{M_K^3}{\Lambda_\chi^3}\, A_i^{\rm scale}
\simeq 0.15\, A_i^{\rm scale},
\end{equation}
and the uncertainty associated with NLO analytic contributions not included in the fits is estimated as
\begin{equation}
\sigma_{{\rm ana},i}
=
\frac{M_K^2}{16\pi^2 f^2}\, A_i^{\rm scale}
\simeq 0.20\, A_i^{\rm scale}.
\end{equation}
The total uncertainty used in the fit is then obtained by adding these contributions in quadrature:
\begin{equation}
\sigma_i
=
\sqrt{\sigma_{i,{\rm input}}^2
+
\sigma_{{\rm trunc},i}^2
+
\sigma_{{\rm ana},i}^2},\label{Eq:uncertainties}
\end{equation}
where $\sigma_{i,{\rm input}}$ denotes the uncertainty shown in Table~\ref{tab:deltaI32_input}. 
To avoid assigning an unrealistically large weight to amplitudes with very small central values, we define
\begin{equation}
\begin{aligned}
A_i^{\rm scale}
&=
\max\left(|A_i^{\rm input}|,A_{\rm floor}\right), \\
A_{\rm floor}
&=
\frac{1}{N_{\rm fit}}
\sum_{i=1}^{N_{\rm fit}} |A_i^{\rm input}|,
\end{aligned}
\end{equation}
with $N_{\rm fit}=4$ for the four $|\Delta I|=3/2$ isospin amplitudes included in the fit.

We perform two sets of fits. In the first, the loop-generated NLO analytic terms are subtracted from the EOMS theoretical expressions, whereas in the second these terms are retained. In both sets, the theoretical expressions are built up progressively to assess the impact of each theoretical component.
The first set compares the EOMS results with the corresponding HB results, which intrinsically contain no NLO analytic contributions. For the HB framework, we adopt the theoretical results obtained in
Ref.~\cite{AbdElHady1999DeltaI32}, except for the pole-diagram
propagator-correction contributions denoted by
$\gamma^{P}_{B_f B_i}$ in Eq.~\eqref{iM_P_structure}, which are
replaced by the corresponding contributions obtained using the treatment
described in
Appendix~\ref{Treatment of propagator corrections in the P-wave pole diagrams}\footnote{
In Ref.~\cite{AbdElHady1999DeltaI32}, $\gamma^{P}_{B_f B_i}$ was treated differently. 
After private communication with one of the authors of Ref.~\cite{AbdElHady1999DeltaI32}, the author agree that the treatment of the $\gamma^{P}_{B_f B_i}$ in their earlier work~\cite{AbdElHady1999DeltaI32} is non-standard, and that our present calculation is appropriate. 
Therefore, when quoting the heavy-baryon results of Ref.~\cite{AbdElHady1999DeltaI32}, 
we replace this particular contribution with the result obtained from the prescription employed here.
}.

The results of the two sets of fits are shown in Tables~\ref{tab:updated_hyperon_amplitudes} and~\ref{tab:analytic_terms_amplitudes}. As discussed above, only the data of four selected amplitudes are taken into account. 
The uncertainties assigned to the fits are obtained from Eq.~(\ref{Eq:uncertainties}).
In addition, because pion-loop contributions have been considered in our fits, only the real parts of the theoretical amplitudes are used to construct the $\chi^2$ function. Notably, the fits performed here are intended only to  assess the impact of different theoretical ingredients, rather than to provide a complete description of the four $|\Delta I|=3/2$ isospin amplitudes.

We use the value of $\chi^2$ as a criterion for the quality of the fits. To visualize how each theoretical component affects the fit quality, we plot in Fig.~\ref{fig:observable_merged} the evolution of $\chi^2$ with respect to each component for the fits listed in Tables~\ref {tab:updated_hyperon_amplitudes} and~\ref{tab:analytic_terms_amplitudes}. 
As shown in Fig.~\ref{fig:observable_merged}, the $\chi^2$ values exhibit an overall downward trend as more theoretical ingredients are included.
More specifically, for the first fit results shown in Fig.~\ref{fig:observable_merged}$(a)$, we first discuss the relativistic effect by comparing the EOMS $\chi$PT results with those from the HB $\chi$PT. Note that the comparison is limited to the O and O+D conventional cases because no corresponding HB results are available for the consistent coupling scheme or for calculations including pion-loop contributions. Here ``O" denotes the inclusion of octet-baryon loop contributions in addition to the tree-level contributions, while ``D" stands for the further inclusion of decuplet-baryon loop contributions, evaluated using the conventional and consistent coupling schemes. In the O case, the $\chi^2$ value decreases from 26.56 in the HB $\chi$PT to 26.31 in the covariant EOMS $\chi$PT. 
In the O+D conventional case, it decreases from 25.36 to 25.13. 
This small reduction in both cases suggests that relativistic corrections have only a limited impact on the fit quality.

One next investigates how the fit quality depends on different theoretical components within the EOMS framework. Fig.~\ref{fig:observable_merged}$(a)$ shows that the $\chi^2$ value decreases from 27.27 to 26.31 when going from the tree-level setup to the O setup. 
The small reduction suggests that octet-baryon loop contributions alone lead to only a mild improvement in the fit quality. One also finds that when going from the O setup to the O+D setup, the $\chi^2$ values depend strongly on the decuplet-coupling scheme. The consistent coupling scheme reduces the $\chi^2$ value from 26.31 to 22.23, whereas the conventional coupling scheme reduces it only to 25.13.
The more pronounced reduction in $\chi^2$ obtained with the consistent coupling scheme suggests that it yields a larger improvement in the fit quality than the conventional one.

In addition, we consider the effect of adding pion-loop contributions to the O+D setup.
In the consistent coupling scheme, these contributions reduce the $\chi^2$ value from 22.23 to 18.63, whereas in the conventional coupling scheme, the reduction is only from 25.13 to 24.85.
The reductions observed in both coupling schemes indicate that pion-loop contributions improve the fit quality, with a more pronounced improvement in the consistent coupling scheme.

For the second set of fit results shown in Fig.~\ref{fig:observable_merged}$(b)$ with the NLO analytic terms retained, the same qualitative pattern as that in Fig.~\ref{fig:observable_merged}$(a)$ is observed. Namely, the consistent coupling scheme still gives a larger reduction in $\chi^2$ than the conventional scheme, and the impact of pion-loop contributions remains more pronounced in the consistent-coupling scheme.
A comparison between Fig.~\ref{fig:observable_merged}$(a)$ and Fig.~\ref{fig:observable_merged}$(b)$ further shows that retaining the NLO loop-generated analytic terms generally reduces the $\chi^2$ values for the considered setups. The only exception is the O setup, where the change is marginal and the $\chi^2$ value remains nearly unchanged.
Overall, while the NLO analytic terms improve the quality of our fits, they do not change the behavior of the $\chi^2$ as a function of the theoretical components.

Finally, we discuss how different theoretical components affect the $|\Delta I|=3/2$ amplitudes. Tables~\ref{tab:updated_hyperon_amplitudes} and~\ref{tab:analytic_terms_amplitudes} show that the inclusion of decuplet-baryon and pion-loop contributions, especially when the consistent coupling scheme is used, tends to bring the fitted amplitude closer to the data. 
Nevertheless, the fitted results still show noticeable deviations from the data, most notably for $P^{\Lambda}_{3,3}$ in both tables.
The remaining discrepancies may arise from the restricted set of loop topologies considered here, the omission of local analytic counterterms, and higher-order contributions. They therefore motivate a more complete analysis of these amplitudes in future work.

\section{summary}\label{Sec4}
In this work, we have reexamined the $|\Delta I|=3/2$ amplitudes of nonleptonic hyperon decays within the  EOMS $\chi$PT. 
The calculation was performed with the same restricted set of diagrammatic topologies as those studied in the early HB $\chi$PT analyses, allowing us to evaluate the relativistic effects. 
We also assess the impact of explicit decuplet-baryon contributions and pion-loop contributions on fit results,
and investigate the dependence of the results on the choice of decuplet-coupling scheme.

The comparison with the HB $\chi$PT results shows that relativistic effects alone lead to only mild changes in the fit quality. 
A more pronounced improvement is obtained when decuplet-baryon contributions are included, especially when the consistent coupling scheme is adopted. 
Pion-loop contributions further reduce the $\chi^2$ values, again most notably in the consistent-coupling scheme. 
These results indicate that the treatment of the decuplet sector is important for describing the selected $|\Delta I|=3/2$ amplitudes. We have also examined the effect of retaining the NLO loop-generated analytic terms.  
One finds that the qualitative pattern of the $\chi^2$
evolution remains the same as that observed in the fits with these terms
subtracted. 
Therefore, the main conclusions are robust to the treatment of the NLO loop-generated analytic terms.

While the inclusion of decuplet-baryon and pion-loop contributions generally improves the description of the $|\Delta I|=3/2$ amplitudes, non-negligible discrepancies persist, particularly in the case of $P^{\Lambda}_{3,3}$. 
These residual discrepancies may arise from the restricted set of loop topologies considered here, the omission of local analytic counterterms, and higher-order chiral contributions. 
A more complete analysis, including the full NLO analytic counterterms and the one-loop topologies not considered here, is needed in future work.

\section{Acknowledgments} 
We thank Profs. Jusak Tandean and Jifeng Yang for useful discussions on the corrections to pole-diagram contributions and the convention differences between relativistic and non-relativistic frameworks. R.X.S acknowledges support from the National Natural Science Foundation of China under Grant No. 12405091, the Natural Science Foundation of Guangxi province under Grant No. 2025GXNSFBA069314, and the Starting Research Fund from the Guangxi Normal University under Grant No.DC2300003299. J.X.L and L.S.G acknowledge support from the National Science Foundation of China under Grant No. W2543006 and Nos. 12435007, 12522505 and 1252200936, and the National Key R\&D Program of China under Grant No. 2023YFA1606703. 
\appendix
\onecolumngrid
\section{Amplitudes}
\label{app:amplitudes}
In this Appendix, we present the loop contributions in integral form, since their fully evaluated expressions are too lengthy to display.
We begin with the contributions involving only octet baryons. 
Before presenting the corresponding loop integrals, we introduce the propagators and vertices used. 
In the following expressions, the algebraic forms are shown on the left, while the corresponding diagrammatic representations are displayed on the right.

For the octet baryon and pseudoscalar meson propagators, we have
{
\renewcommand{\arraystretch}{1.20}

\begin{center}
\hspace*{-0.0\linewidth}
\begin{tabular}{@{}c@{\hspace{3em}}c@{}}

$\displaystyle
S_B(p) =
\frac{i(\slashed{p}+m_b)}{p^2-m_b^2+i\epsilon},
$
&
\raisebox{-0.2\height}{\includegraphics[width=0.15\linewidth]{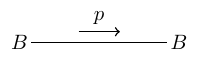}}
\\[-0.4em]

$\displaystyle
\Delta_\phi(l) =
\frac{i}{l^2-m_\phi^2+i\epsilon}.
$
&
\raisebox{-0.33\height}{\includegraphics[width=0.15\linewidth]{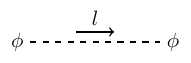}}

\end{tabular}
\end{center}

}

The interaction vertices derived from the corresponding Lagrangians are listed below.
For compactness, we display only their Dirac and Lorentz structures, omitting the corresponding flavor factors and overall coefficients.

The weak vertex is given by
{
\renewcommand{\arraystretch}{1.20}

\begin{center}
\begin{tabular}{@{}c@{\hspace{3em}}c@{}}
$\displaystyle \mathcal{V}_{\rm w}=i,$
&
\raisebox{-0.33\height}{\includegraphics[width=0.12\linewidth]{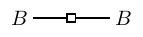}}
\end{tabular}
\end{center}

}

The same weak vertex structure is used when additional meson lines are attached to the weak vertex, and such vertices are therefore not displayed separately.

For the strong meson--octet--baryon vertex, we have
{
\renewcommand{\arraystretch}{1.20}

\begin{center}
\begin{tabular}{@{}c@{\hspace{0.5em}}c@{}}
$\displaystyle
\mathcal{V}_{B\phi B}(q)
=
i\,\slashed{q}\gamma^5,
$
&
\raisebox{-0.15\height}{
  \includegraphics[width=0.13\linewidth]{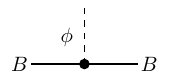}
}
\end{tabular}
\end{center}

}

For the strong vertices involving two meson lines, the two Dirac structures needed in the loop integrals are
{
\renewcommand{\arraystretch}{1.20}

\begin{center}
\hspace*{-0.05\linewidth}
\begin{tabular}{@{}c@{\hspace{0.6em}}c@{}}

$\displaystyle
\mathcal{V}_{B\phi\phi B}^{(+)}(q_1,q_2)
=
i(\slashed{q}_1+\slashed{q}_2),
$
&
\raisebox{-0.43\height}{\includegraphics[width=0.13\linewidth]{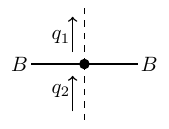}}
\\[-0.2em]

$\displaystyle
\mathcal{V}_{B\phi\phi B}^{(-)}(q_1,q_2)
=
i(\slashed{q}_1-\slashed{q}_2).
$
&
\raisebox{-0.43\height}{\includegraphics[width=0.13\linewidth]{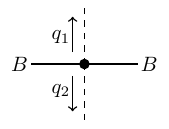}}

\end{tabular}
\end{center}

}
The relative sign between the $\slashed q_1$ and $\slashed q_2$ terms is fixed by the relative momentum flow of the two meson lines at the vertex. 
It is therefore part of the vertex structure itself, rather than an overall sign of the diagram, and must be kept explicitly in $\mathcal{V}_{B\phi\phi B}^{(+)}$ and $\mathcal{V}_{B\phi\phi B}^{(-)}$. 

Using the vertices and propagators introduced above, the amplitudes $\Sigma_{ij}^{x}$ associated with the octet-baryon loop diagrams in Figs.~\ref{fig:loop_swave} and~\ref{fig:loop_pwave} are given by
\begin{equation}
\Sigma_{ij}^{c}
=
\beta_{ij}^{c}
\int\!\frac{\mu^{4-D}\,d^Dl}{(2\pi)^D}\,
\bar u(p_f)\,
\mathcal{V}_w\,
u(p_i)\,
\Delta_\phi(l).
\end{equation}

\vspace{-0.5cm}
\begin{align}
\Sigma_{ij}^{d_1}
={}&
\beta_{ij}^{d_1}
\int\!\frac{\mu^{4-D}\,d^Dl}{(2\pi)^D}\,
\bar u(p_f)\,
\mathcal{V}_{B\phi\phi B}^{(+)}(k,l)\,
S_B(p_i-l)\,
\mathcal{V}_w\,
u(p_i)\,\\
&\times\Delta_\phi(l).
\end{align}

\vspace{-0.5cm}
\begin{align}
\Sigma_{ij}^{d_2}
={}&
\beta_{ij}^{d_2}
\int\!\frac{\mu^{4-D}\,d^Dl}{(2\pi)^D}\,
\bar u(p_f)\,
\mathcal{V}_w\,
S_B(p_f-l)\,
\mathcal{V}_{B\phi\phi B}^{(-)}(k,l)\,
u(p_i)\,\\
&\times\Delta_\phi(l).
\end{align}

\vspace{-0.5cm}
\begin{align}
\Sigma_{ij}^{e}
={}&
\beta_{ij}^{e}
\int\!\frac{\mu^{4-D}\,d^Dl}{(2\pi)^D}\,
\bar u(p_f)\,
\mathcal{V}_{B\phi B}(l)\,
S_B(p_f-l)\,
\mathcal{V}_w\,\\&\times
S_B(p_i-l)\,
\mathcal{V}_{B\phi B}(l)\,
u(p_i)\,
\Delta_\phi(l).
\end{align}

\vspace{-0.5cm}
\begin{align}
\Sigma_{ij}^{g_1}
={}&
\beta_{ij}^{g_1}
\int\!\frac{\mu^{4-D}\,d^Dl}{(2\pi)^D}\,
\bar u(p_f)\,
\mathcal{V}_{B\phi B}(k)\,
S_B(p_i)\,
\mathcal{V}_w\,
u(p_i)\,
\Delta_\phi(l).
\end{align}

\vspace{-0.5cm}
\begin{align}
\Sigma_{ij}^{g_2}
={}&
\beta_{ij}^{g_2}
\int\!\frac{\mu^{4-D}\,d^Dl}{(2\pi)^D}\,
\bar u(p_f)\,
\mathcal{V}_w\,
S_B(p_f)\,
\mathcal{V}_{B\phi B}(k)\,
u(p_i)\,
\Delta_\phi(l).
\end{align}

\vspace{-0.5cm}
\begin{align}
\Sigma_{ij}^{h_1}
={}&
\beta_{ij}^{h_1}
\int\!\frac{\mu^{4-D}\,d^Dl}{(2\pi)^D}\,
\bar u(p_f)\,
\mathcal{V}_{B\phi B}(k)\,
S_B(p_i)\,
\mathcal{V}_w\,
u(p_i)\,
\Delta_\phi(l).
\end{align}

\vspace{-0.5cm}
\begin{align}
\Sigma_{ij}^{h_2}
={}&
\beta_{ij}^{h_2}
\int\!\frac{\mu^{4-D}\,d^Dl}{(2\pi)^D}\,
\bar u(p_f)\,
\mathcal{V}_w\,
S_B(p_f)\,
\mathcal{V}_{B\phi B}(k)\,
u(p_i)\,
\Delta_\phi(l).
\end{align}
\begin{align}
\Sigma_{ij}^{i_1}
={}&
\beta_{ij}^{i_1}
\int\!\frac{\mu^{4-D}\,d^Dl}{(2\pi)^D}\,
\bar u(p_f)\,
\mathcal{V}_{B\phi B}(l)\,
S_B(p_f-l)\,
\mathcal{V}_{B\phi B}(k)\,
\nonumber\\
&\times
S_B(p_i-l)\,
\mathcal{V}_{B\phi B}(l)\,
S_B(p_i)\,
\mathcal{V}_w\,
u(p_i)\,
\Delta_\phi(l).
\end{align}

\vspace{-0.5cm}
\begin{align}
\Sigma_{ij}^{i_2}
={}&
\beta_{ij}^{i_2}
\int\!\frac{\mu^{4-D}\,d^Dl}{(2\pi)^D}\,
\bar u(p_f)\,
\mathcal{V}_w\,
S_B(p_f)\,
\mathcal{V}_{B\phi B}(l)\,
\nonumber\\
&\times
S_B(p_f-l)\,
\mathcal{V}_{B\phi B}(k)\,
S_B(p_i-l)\,
\mathcal{V}_{B\phi B}(l)\,
u(p_i)\,
\Delta_\phi(l).
\end{align}

\vspace{-0.5cm}
\begin{align}
\Sigma_{ij}^{m_1}
={}&
\beta_{ij}^{m_1}
\int\!\frac{\mu^{4-D}\,d^Dl}{(2\pi)^D}\,
\bar u(p_f)\,
\mathcal{V}_{B\phi B}(k)\,
S_B(p_i)\,
\mathcal{V}_{B\phi B}(l)\,
\nonumber\\
&\times
S_B(p_i-l)\,
\mathcal{V}_w\,
S_B(p_i-l)\,
\mathcal{V}_{B\phi B}(l)\,
u(p_i)\,
\Delta_\phi(l).
\end{align}

\vspace{-0.5cm}
\begin{align}
\Sigma_{ij}^{m_2}
={}&
\beta_{ij}^{m_2}
\int\!\frac{\mu^{4-D}\,d^Dl}{(2\pi)^D}\,
\bar u(p_f)\,
\mathcal{V}_{B\phi B}(l)\,
S_B(p_f-l)\,
\mathcal{V}_w\,
\nonumber\\
&\times
S_B(p_f-l)\,
\mathcal{V}_{B\phi B}(l)\,
S_B(p_f)\,
\mathcal{V}_{B\phi B}(k)\,
u(p_i)\,
\Delta_\phi(l).
\end{align}
The amplitudes $\Sigma_{ij}^{\prime x}$ denote the contributions from the loop diagrams involving intermediate decuplet baryons labeled by $x$ in Figs.~\ref{fig:loop_swave} and~\ref{fig:loop_pwave}. 
The coefficient $\beta^{x}_{ij}$ contains the corresponding flavor structure and, as discussed above, also absorbs the relevant coupling constants and simple numerical prefactors, such as overall signs, in order to keep the expressions compact.

We next discuss the loop contributions involving decuplet baryons. 
The propagator and interaction vertices needed for these contributions are collected below.

The decuplet baryon is described by the Rarita--Schwinger propagator,

{
\renewcommand{\arraystretch}{1.20}

\begin{center}
\hspace*{-0.075\linewidth}
\begin{tabular}{@{}c@{\hspace{3em}}c@{}}

$\displaystyle
S_{\mu\nu}(p)
=
\frac{i(\slashed{p}+M_D)}{p^2-M_D^2+i\epsilon}
\,\mathcal{P}_{\mu\nu}^{(3/2)}(p)
,
$
&
\raisebox{-0.33\height}{\includegraphics[width=0.15\linewidth]{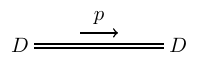}}

\end{tabular}
\end{center}

}

\begin{equation*}
\begin{aligned}
\mathcal{P}_{\mu\nu}^{(3/2)}(p)
=
&-\,g_{\mu\nu}
+\frac{1}{D-1}\gamma_\mu\gamma_\nu
+\frac{1}{(D-1)M_D}\left(\gamma_\mu p_\nu-\gamma_\nu p_\mu\right)
+\frac{D-2}{(D-1)M_D^2}p_\mu p_\nu.
\end{aligned}
\end{equation*}

The following structures represent the weak interaction vertices involving decuplet baryons:
{
\renewcommand{\arraystretch}{1.20}

\begin{center}
\begin{tabular}{@{}c@{\hspace{1.1em}}c@{}}

$\displaystyle
\mathcal{V}_{\rm w}^{\mu\nu}
=
i\,g^{\mu\nu}.
$
&
\raisebox{-0.33\height}{\includegraphics[width=0.14\linewidth]{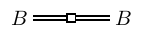}}

\end{tabular}
\end{center}

}

For the strong vertices involving decuplet baryons, two different coupling schemes are considered. 
The corresponding vertex structures are given below.

The decuplet--meson--octet baryon vertex has the following consistent- and conventional-coupling structures:
{
\renewcommand{\arraystretch}{1.20}

\begin{center}
\hspace*{-0.05\linewidth}
\begin{tabular}{@{}c@{\hspace{1.2em}}c@{\hspace{4em}}c@{}}

$\displaystyle
\mathcal{V}_{D\phi B}^{\mu}(p_D,q)
=
\frac{i}{M_D}\,
p_{D,\alpha}\,q_\nu\,\gamma^{\alpha\mu\nu},
$
&
$\displaystyle
\mathcal{V}_{D\phi B,\mathrm{conv}}^{\mu}(q)
=
i\,q^\mu.
$
&
\raisebox{-0.17\height}{\includegraphics[width=0.13\linewidth]{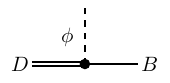}}

\end{tabular}
\end{center}

}
Here, the subscript ``conv'' denotes the conventional-coupling scheme, while the absence of this subscript indicates the consistent-coupling scheme. 

For the decuplet--meson--decuplet vertex, the corresponding consistent- and conventional-coupling structures are
{
\renewcommand{\arraystretch}{1.20}

\begin{center}
\hspace*{-0.05\linewidth}
\begin{tabular}{@{}c@{\hspace{1.2em}}c@{\hspace{1.2em}}c@{}}

$\displaystyle
\mathcal{V}_{D\phi D}^{\mu\nu}(p_D,q)
=
\frac{i}{M_D}\,
p_{D,\rho}\,q_\sigma\,\gamma^{\mu\nu\rho\sigma}\gamma^5,
$
&
$\displaystyle
\mathcal{V}_{D\phi D,\mathrm{conv}}^{\mu\nu}(q)
=
i\,g^{\mu\nu}\slashed{q}\gamma^5.
$
&
\raisebox{-0.17\height}{\includegraphics[width=0.13\linewidth]{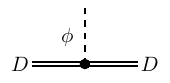}}

\end{tabular}
\end{center}

}

With the vertices and propagators specified above, the amplitudes $\Sigma_{ij}^{x}$ for the decuplet-baryon loop diagrams labeled by $x$ in Fig.~\ref{fig:loop_pwave} are written as
\vspace{-0.3cm}
\begin{align}
\Sigma_{ij}^{\prime f}
={}&
\beta_{ij}^{\prime f}
\int\!\frac{\mu^{4-D}\,d^Dl}{(2\pi)^D}\,
\bar u(p_f)\,
\mathcal{V}_{D\phi B}^{\mu}(p_f-l,l)\,
S_{\mu\nu}(p_f-l)\,
\nonumber\\
&\times
\mathcal{V}_w^{\nu\lambda}\,
S_{\lambda\sigma}(p_i-l)\,
\mathcal{V}_{D\phi B}^{\sigma}(p_i-l,l)\,
u(p_i)\,
\Delta_\phi(l).
\end{align}

\vspace{-0.3cm}
\begin{align}
\Sigma_{ij}^{\prime j_1}
={}&
\beta_{ij}^{j_1}
\int\!\frac{\mu^{4-D}\,d^Dl}{(2\pi)^D}\,
\bar u(p_f)\,
\mathcal{V}_{D\phi B}^{\mu}(p_f-l,l)\,
S_{\mu\nu}(p_f-l)\,
\mathcal{V}_{D\phi D}^{\nu\lambda}(p_i-l,k)\,
\nonumber\\
&\times
S_{\lambda\sigma}(p_i-l)\,
\mathcal{V}_{D\phi B}^{\sigma}(p_i-l,l)\,
\Delta_\phi(l)\,
S_B(p_i)\,
\mathcal{V}_w\,
u(p_i).
\end{align}

\vspace{-0.3cm}
\begin{align}
\Sigma_{ij}^{\prime j_2}
={}&
\beta_{ij}^{j_2}
\int\!\frac{\mu^{4-D}\,d^Dl}{(2\pi)^D}\,
\bar u(p_f)\,
\mathcal{V}_w\,
S_B(p_f)\,
\mathcal{V}_{D\phi B}^{\mu}(p_f-l,l)\,
S_{\mu\nu}(p_f-l)
\nonumber\\
&\times
\mathcal{V}_{D\phi D}^{\nu\lambda}(p_i-l,k)\,
S_{\lambda\sigma}(p_i-l)\,
\mathcal{V}_{D\phi B}^{\sigma}(p_i-l,l)\,
\Delta_\phi(l)\,
u(p_i).
\end{align}

\vspace{-0.3cm}
\begin{align}
\Sigma_{ij}^{\prime k_1}
={}&
\beta_{ij}^{k_1}
\int\!\frac{\mu^{4-D}\,d^Dl}{(2\pi)^D}\,
\bar u(p_f)\,
\mathcal{V}_{B\phi B}(l)\,
S_B(p_f-l)\,
\mathcal{V}_{D\phi B}^{\mu}(p_i-l,k)\,\nonumber\\
&\times
S_{\mu\nu}(p_i-l)
\mathcal{V}_{D\phi B}^{\nu}(p_i-l,l)\,
\Delta_\phi(l)\,
S_B(p_i)\,
\mathcal{V}_w\,
u(p_i).
\end{align}

\vspace{-0.3cm}
\begin{align}
\Sigma_{ij}^{\prime k_2}
={}&
\beta_{ij}^{k_2}
\int\!\frac{\mu^{4-D}\,d^Dl}{(2\pi)^D}\,
\bar u(p_f)\,
\mathcal{V}_w\,
S_B(p_f)\,
\mathcal{V}_{B\phi B}(l)\,
S_B(p_f-l)\,
\nonumber\\
&\times
\mathcal{V}_{D\phi B}^{\mu}(p_i-l,k)\,
S_{\mu\nu}(p_i-l)\,
\mathcal{V}_{D\phi B}^{\nu}(p_i-l,l)\,
\Delta_\phi(l)\,
u(p_i).
\end{align}

\vspace{-0.3cm}
\begin{align}
\Sigma_{ij}^{\prime l_1}
={}&
\beta_{ij}^{l_1}
\int\!\frac{\mu^{4-D}\,d^Dl}{(2\pi)^D}\,
\bar u(p_f)\,
\mathcal{V}_{D\phi B}^{\mu}(p_f-l,l)\,
S_{\mu\nu}(p_f-l)\,
\mathcal{V}_{D\phi B}^{\nu}(p_f-l,k)\,\nonumber\\
&\times
S_B(p_i-l)
\mathcal{V}_{B\phi B}(l)\,
\Delta_\phi(l)\,
S_B(p_i)\,
\mathcal{V}_w\,
u(p_i).
\end{align}

\vspace{-0.3cm}
\begin{align}
\Sigma_{ij}^{\prime l_2}
={}&
\beta_{ij}^{l_2}
\int\!\frac{\mu^{4-D}\,d^Dl}{(2\pi)^D}\,
\bar u(p_f)\,
\mathcal{V}_w\,
S_B(p_f)\,
\mathcal{V}_{D\phi B}^{\mu}(p_f-l,l)\,
S_{\mu\nu}(p_f-l)\,
\nonumber\\
&\times
\mathcal{V}_{D\phi B}^{\nu}(p_f-l,k)\,
S_B(p_i-l)\,
\mathcal{V}_{B\phi B}(l)\,
\Delta_\phi(l)\,
u(p_i).
\end{align}

\vspace{-0.3cm}
\begin{align}
\Sigma_{ij}^{\prime n_1}
={}&
\beta_{ij}^{n_1}
\int\!\frac{\mu^{4-D}\,d^Dl}{(2\pi)^D}\,
\bar u(p_f)\,
\mathcal{V}_{B\phi B}(k)\,
S_B(p_i)\,
\mathcal{V}_{D\phi B}^{\mu}(p_i-l,l)\,
\nonumber\\
&\times S_{\mu\nu}(p_i-l)\,
\mathcal{V}_w^{\nu\lambda}\,
S_{\lambda\sigma}(p_i-l)\,
\mathcal{V}_{D\phi B}^{\sigma}(p_i-l,l)\,
\Delta_\phi(l)\,
u(p_i).
\end{align}

\vspace{-0.3cm}
\begin{align}
\Sigma_{ij}^{\prime n_2}
={}&
\beta_{ij}^{n_2}
\int\!\frac{\mu^{4-D}\,d^Dl}{(2\pi)^D}\,
\bar u(p_f)\,
\mathcal{V}_{D\phi B}^{\mu}(p_f-l,l)\,
S_{\mu\nu}(p_f-l)\,
\mathcal{V}_w^{\nu\lambda}\,
\nonumber\\
&\times
S_{\lambda\sigma}(p_f-l)\,
\mathcal{V}_{D\phi B}^{\sigma}(p_f-l,l)\,
\Delta_\phi(l)\,
S_B(p_f)\,
\mathcal{V}_{B\phi B}(k)\,
u(p_i).
\end{align}

The amplitudes $\Sigma_{ij}^{\prime x}$ denote the contributions from the loop diagrams involving intermediate decuplet baryons.
The loop integrals above are written for the consistent-coupling scheme. 
The corresponding conventional-coupling expressions are obtained by replacing the consistent vertices in the integrands with the conventional ones, and are therefore not displayed separately.
After evaluating the loop integrals given above, the resulting expressions correspond to
those obtained in the heavy-baryon formalism in
Ref.~\cite{AbdElHady1999DeltaI32}.

\section{Wave-function renormalization factors and decay-constant correction}
\label{app:wfr}

In this section, we collect the wave-function renormalization factors and the
decay-constant correction used in Eq.~\eqref{eq:iM_structure}. We adopt the convention
\begin{equation}
F_\pi=F_{0}\left(1+\delta F_\pi\right),
\qquad
\frac{1}{F_{0}}
=
\frac{1}{F_\pi}(1+\delta F_\pi),
\end{equation}
so that the replacement of the pseudoscalar decay constant in the chiral limit
by the physical pion decay constant gives the contribution $\delta F_\pi$ to
$\lambda_{B_fB_i\pi}$.

The one-point function is defined as
\begin{equation}
I(M^2)
=
-\frac{M^2}{16\pi^2}
\ln\frac{M^2}{\mu^2}.
\end{equation}
With this convention, the pion decay-constant correction up to NLO reads~\cite{BorasoyHolstein1999,Lu2019MesonBaryon}
\begin{equation}
\delta F_\pi
=
\frac{1}{F_\pi^2}
\left[
4L_4\left(2m_K^2+m_\pi^2\right)
+
4L_5m_\pi^2
+
I(m_\pi^2)
+
\frac{1}{2}I(m_K^2)
\right].
\end{equation}
The pion wave-function renormalization factor up to NLO is
\begin{equation}
Z_\pi
=
1-\frac{1}{F_\pi^2}
\left[
8L_4\left(2m_K^2+m_\pi^2\right)
+
8L_5m_\pi^2
+
\frac{1}{3}I(m_K^2)
+
\frac{2}{3}I(m_\pi^2)
\right].
\end{equation}
In the numerical analysis, the LECs $L_4$ and $L_5$ appearing in
$Z_\pi$ and $\delta F_\pi$ are set to zero.

\begin{figure}
    \centering
    \includegraphics[width=0.4\linewidth]{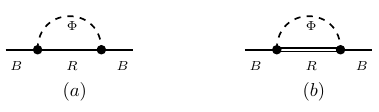}
    \caption{Wave-function renormalization contributions to octet-baryon fields.}
    \label{fig:wfr_octet_baryons}
\end{figure}

We next give the octet-baryon wave-function renormalization factors. The relevant self-energy diagrams are shown in Fig.~\ref{fig:wfr_octet_baryons}. The contribution generated by an intermediate meson--baryon state $(\Phi,R)$ can be decomposed as~\cite{Scherer2003ChPT,Lu2019MesonBaryon}
\begin{equation}
\Sigma_X(B,\Phi,R;\slashed p)
=
-
\widehat f_X(s;m_R,M_\Phi)\slashed p
+
\widehat g_X(s;m_R,M_\Phi)m_R ,
\qquad
X=O,D ,
\label{eq:self_energy_decomposition_wfr}
\end{equation}
where $X=O$ and $X=D$ denote intermediate octet and decuplet baryons,
respectively. Here, $m_R$ and $m_\Phi$ denote the masses of the intermediate baryon and meson, respectively, $p$ is the momentum of the external baryon $B$, and $s=p^2$ is the corresponding invariant momentum squared. The index specifying the decuplet-coupling scheme is suppressed.

For the specified external baryon $B$, intermediate meson $\Phi$, and
intermediate baryon $R$, the corresponding single-channel contribution to the
wave-function renormalization factor is written as
\begin{equation}
\delta Z_X(B,\Phi,R)
=
-
\left[
f_X(B,\Phi,R)
+
2m_B^2 f_X'(B,\Phi,R)
-
2m_Bm_R g_X'(B,\Phi,R)
\right]
-
\delta Z_X^{\rm PCB},
\label{eq:deltaZ_unified}
\end{equation}
where all quantities are evaluated on the mass shell $s=m_B^2$. Here the
prime denotes the derivative with respect to $s=p^2$,
\begin{align}
f_X(B,\Phi,R)
&=
\left.
\widehat f_X(s;m_R,M_\Phi)
\right|_{s=m_B^2},
&
f_X'(B,\Phi,R)
&=
\left.
\frac{\partial}{\partial s}
\widehat f_X(s;m_R,M_\Phi)
\right|_{s=m_B^2},
\\
g_X(B,\Phi,R)
&=
\left.
\widehat g_X(s;m_R,M_\Phi)
\right|_{s=m_B^2},
&
g_X'(B,\Phi,R)
&=
\left.
\frac{\partial}{\partial s}
\widehat g_X(s;m_R,M_\Phi)
\right|_{s=m_B^2}.
\end{align}
The EOMS subtraction term in Eq.~\eqref{eq:deltaZ_unified} depends on the
intermediate baryon multiplet. For intermediate octet baryons,
\begin{equation}
\delta Z_O^{\rm PCB}=0,
\label{eq:deltaZO_PCB}
\end{equation}
whereas for intermediate decuplet baryons, the subtraction term takes the same
form in both coupling schemes,
\begin{equation}
\delta Z_D^{\rm PCB}
=
\frac{m_0}{1152\pi^2}
\left[
-30(5m_0-6m_D)\ln\frac{m_0^2}{\mu^2}
+85m_0
-78m_D
\right].
\label{eq:deltaZD_PCB}
\end{equation}

For intermediate octet baryons, the scalar functions entering
Eq.~\eqref{eq:deltaZ_unified} are
\begin{align}
\widehat f_O(s;m_R,M_\Phi)
&=
\frac{1}{32\pi^2s}
\Big[
\left(
-m_R^2(2s+M_\Phi^2)
+s(s-M_\Phi^2)
+m_R^4
\right)
B_0(s,m_R^2,M_\Phi^2)
\nonumber\\
&
+
(m_R^2-s)A_0(M_\Phi^2)
-
(s+m_R^2)A_0(m_R^2)
\Big],
\\
\widehat g_O(s;m_R,M_\Phi)
&=
\frac{1}{16\pi^2}
\left[
M_\Phi^2B_0(s,m_R^2,M_\Phi^2)
+
A_0(m_R^2)
\right].
\end{align}
Here \(A_0\) and \(B_0\) are the one- and two-point scalar functions in the Passarino-Veltman notation~\cite{PassarinoVeltman1979OneLoopCorrections}. In the numerical evaluation, we set
\(m_R=m_0\) for the intermediate octet baryons.

We obtain the octet contribution to the wave-function renormalization factors by summing the contributions from all allowed intermediate states, weighted by their respective flavor factors.
Explicitly,
\begin{align}
\Delta Z_N^{(O)}
&=
\frac{1}{12F_\pi^2}
\Big[
(D-3F)^2 \delta Z_O(N,\eta,N)
+
(D+3F)^2 \delta Z_O(N,K,\Lambda)
\nonumber\\
&\hspace{2.3cm}
+
9(D-F)^2 \delta Z_O(N,K,\Sigma)
+
9(D+F)^2 \delta Z_O(N,\pi,N)
\Big],
\\[0.5em]
\Delta Z_\Sigma^{(O)}
&=
\frac{1}{6F_\pi^2}
\Big[
3(D-F)^2 \delta Z_O(\Sigma,K,N)
+
3(D+F)^2 \delta Z_O(\Sigma,K,\Xi)
\nonumber\\
&\hspace{2.3cm}
+
2D^2\delta Z_O(\Sigma,\eta,\Sigma)
+
2D^2\delta Z_O(\Sigma,\pi,\Lambda)
+
12F^2\delta Z_O(\Sigma,\pi,\Sigma)
\Big],
\\[0.5em]
\Delta Z_\Lambda^{(O)}
&=
\frac{1}{6F_\pi^2}
\Big[
2D^2\delta Z_O(\Lambda,\eta,\Lambda)
+
6D^2\delta Z_O(\Lambda,\pi,\Sigma)
+
(D-3F)^2 \delta Z_O(\Lambda,K,\Xi)
\nonumber\\
&\hspace{2.3cm}
+
(D+3F)^2 \delta Z_O(\Lambda,K,N)
\Big],
\\[0.5em]
\Delta Z_\Xi^{(O)}
&=
\frac{1}{12F_\pi^2}
\Big[
(D-3F)^2 \delta Z_O(\Xi,K,\Lambda)
+
(D+3F)^2 \delta Z_O(\Xi,\eta,\Xi)
\nonumber\\
&\hspace{2.3cm}
+
9(D-F)^2 \delta Z_O(\Xi,\pi,\Xi)
+
9(D+F)^2 \delta Z_O(\Xi,K,\Sigma)
\Big].
\end{align}

For intermediate decuplet baryons, the scalar functions
$\widehat f_D$ and $\widehat g_D$ depend on the decuplet coupling scheme.
For compactness, we introduce
\begin{equation}
B_{R\Phi}(s)
\equiv
B_0(s,m_R^2,M_\Phi^2),
\qquad
A_R
\equiv
A_0(m_R^2),
\qquad
A_\Phi
\equiv
A_0(M_\Phi^2).
\end{equation}

For the consistent coupling scheme, the scalar functions before setting
$s=m_B^2$ are
\begin{align}
\widehat f_D^{\rm cons}(s;m_R,M_\Phi)
&=
\frac{1}{1152\pi^2m_D^2s}
\Big\{
\Big[
-6s^3
+6s^2m_R^2
+18s^2M_\Phi^2
-6m_R^6
+18m_R^4M_\Phi^2
\nonumber\\
&\hspace{1.7cm}
+6sm_R^4
-18m_R^2M_\Phi^4
+12sm_R^2M_\Phi^2
+6M_\Phi^6
-18sM_\Phi^4
\Big]B_{R\Phi}(s)
\nonumber\\
&\hspace{1.7cm}
+6A_\Phi
\left(
2m_R^2M_\Phi^2
-m_R^4
+3sM_\Phi^2
-M_\Phi^4
+s^2
\right)
\nonumber\\
&\hspace{1.7cm}
+6A_R
\left[
-m_R^2(2M_\Phi^2+s)
+m_R^4
+(s-M_\Phi^2)^2
\right]
\nonumber\\
&\hspace{1.7cm}
-4s^2m_R^2
+3sm_R^4
-8s^2M_\Phi^2
-3sM_\Phi^4
+2s^3
\Big\},
\\[0.5em]
\widehat g_D^{\rm cons}(s;m_R,M_\Phi)
&=
-\frac{1}{288\pi^2m_D^2}
\Big\{
\Big[
-3s^2
-3m_R^4
+6m_R^2M_\Phi^2
+6sm_R^2
-3M_\Phi^4
+6sM_\Phi^2
\Big]B_{R\Phi}(s)
\nonumber\\
&\hspace{1.7cm}
+3A_R(m_R^2-M_\Phi^2+s)
+3A_\Phi(-m_R^2+M_\Phi^2+s)
\nonumber\\
&\hspace{1.7cm}
-3sm_R^2
-3sM_\Phi^2
+s^2
\Big\}.
\end{align}

For the conventional coupling scheme, the corresponding functions are
\begin{align}
\widehat f_D^{\rm conv}(s;m_R,M_\Phi)
&=
\frac{1}{1152\pi^2m_D^2s}
\Big\{
\Big[
24m_D^2m_R^2M_\Phi^2
-24m_D^2M_\Phi^4
+24m_D^2sM_\Phi^2
\nonumber\\
&\hspace{1.7cm}
-6s^3
+6s^2m_R^2
+18s^2M_\Phi^2
-6m_R^6
-6m_R^4M_\Phi^2
+6sm_R^4
\nonumber\\
&\hspace{1.7cm}
+6m_R^2M_\Phi^4
-12sm_R^2M_\Phi^2
+6M_\Phi^6
-18sM_\Phi^4
\Big]B_{R\Phi}(s)
\nonumber\\
&\hspace{1.7cm}
+6A_\Phi
\left[
-2m_R^2M_\Phi^2
-m_R^4
+M_\Phi^2(4m_D^2-5s)
-M_\Phi^4
+s^2
\right]
\nonumber\\
&\hspace{1.7cm}
+6A_R
\left[
-m_R^2(s-2M_\Phi^2)
+m_R^4
-2M_\Phi^2(2m_D^2+s)
+M_\Phi^4
+s^2
\right]
\nonumber\\
&\hspace{1.7cm}
+8sm_R^2M_\Phi^2
-4s^2m_R^2
+3sm_R^4
-8m_D^2sM_\Phi^2
-8s^2M_\Phi^2
\nonumber\\
&\hspace{1.7cm}
+13sM_\Phi^4
+2s^3
\Big\},
\\[0.5em]
\widehat g_D^{\rm conv}(s;m_R,M_\Phi)
&=
-\frac{1}{288\pi^2m_D^2}
\Big\{
\Big[
12m_D^2M_\Phi^2
-3s^2
-3m_R^4
-6m_R^2M_\Phi^2
+6sm_R^2
\nonumber\\
&\hspace{1.7cm}
-3M_\Phi^4
+6sM_\Phi^2
\Big]B_{R\Phi}(s)
\nonumber\\
&\hspace{1.7cm}
+3A_R(-3m_R^2-M_\Phi^2+4m_D^2+s)
+3A_\Phi(-m_R^2-3M_\Phi^2+s)
\nonumber\\
&\hspace{1.7cm}
+4m_R^2M_\Phi^2
-4m_D^2m_R^2
-3sm_R^2
+4m_R^4
-4m_D^2M_\Phi^2
\nonumber\\
&\hspace{1.7cm}
-3sM_\Phi^2
+4M_\Phi^4
+s^2
\Big\}.
\end{align}
In the numerical evaluation of the decuplet-intermediate-state contribution,
the intermediate-baryon mass entering these scalar functions is set to the decuplet mass in the chiral limit, $m_R=m_D$.

With these definitions, the decuplet-intermediate-state contributions to the
octet-baryon wave-function renormalization factors have the same channel-summed
form in the two coupling schemes. Explicitly,
\begin{align}
\Delta Z_N^{(D)}
&=
-\frac{\mathcal C^2}{F_\pi^2}
\left[
\delta Z_D(N,\pi,\Delta)
+
\frac{1}{4}\delta Z_D(N,K,\Sigma^*)
\right],
\\[0.5em]
\Delta Z_\Sigma^{(D)}
&=
-\frac{\mathcal C^2}{F_\pi^2}
\left[
\frac{2}{3}\delta Z_D(\Sigma,K,\Delta)
+
\frac{1}{6}\delta Z_D(\Sigma,\pi,\Sigma^*)
+
\frac{1}{4}\delta Z_D(\Sigma,\eta,\Sigma^*)
+
\frac{1}{6}\delta Z_D(\Sigma,K,\Xi^*)
\right],
\\[0.5em]
\Delta Z_\Lambda^{(D)}
&=
-\frac{\mathcal C^2}{F_\pi^2}
\left[
\frac{3}{4}\delta Z_D(\Lambda,\pi,\Sigma^*)
+
\frac{1}{2}\delta Z_D(\Lambda,K,\Xi^*)
\right],
\\[0.5em]
\Delta Z_\Xi^{(D)}
&=
-\frac{\mathcal C^2}{F_\pi^2}
\left[
\frac{1}{4}\delta Z_D(\Xi,K,\Sigma^*)
+
\frac{1}{4}\delta Z_D(\Xi,\pi,\Xi^*)
+
\frac{1}{4}\delta Z_D(\Xi,\eta,\Xi^*)
+
\frac{1}{2}\delta Z_D(\Xi,K,\Omega)
\right].
\end{align}

Combining the octet- and decuplet-intermediate-state contributions, the full
octet-baryon wave-function renormalization factor is written as
\begin{equation}
Z_B
=
1
+
\Delta Z_B^{(O)}
+
\Delta Z_B^{(D)} .
\end{equation}

Combining these ingredients, the factor multiplying the tree-level amplitude is
\begin{equation}
\lambda_{B_fB_i\pi}
=
\frac{1}{2}\left(Z_{B_f}-1\right)
+
\frac{1}{2}\left(Z_{B_i}-1\right)
+
\frac{1}{2}\left(Z_\pi-1\right)
+
\delta F_\pi .
\end{equation}
The wave-function renormalization factors given above correspond to those
obtained in the heavy-baryon formalism in Refs.~\cite{Jenkins1992,Jenkins1992Mass,AbdElHady1999DeltaI32}.

\section{Treatment of propagator corrections in the $P$-wave pole diagrams}
\label{Treatment of propagator corrections in the P-wave pole diagrams}

In the $P$-wave amplitudes, the pole diagrams shown in Fig.~\ref{fig:tree_diagrams} $(b1)$ and $(b2)$ contain intermediate baryon propagators. 
The corresponding one-loop self-energy insertions on these propagators are shown in Fig.~\ref{fig:loop_extra_pwave}. Following the treatment in Refs.~\cite{Springer1995Revisited,BorasoyHolstein1999}, these diagrams are not added as independent loop amplitudes.
In this section, we describe how these propagator corrections are treated in the present calculation.

\begin{figure}[htbp]
\centering

\includegraphics[width=0.8\textwidth]{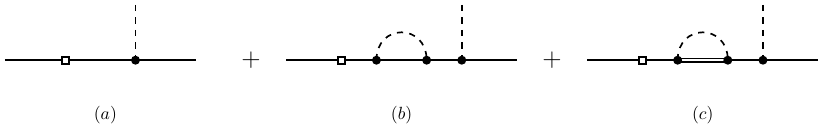}

\caption{
Schematic illustration of a $P$-wave pole diagram and the corresponding one-loop self-energy insertion on the intermediate baryon propagator.
}
\label{fig:pwave_pole_selfenergy}
\end{figure}

First, we consider a $P$-wave pole diagram with one-loop self-energy insertions in the intermediate baryon propagator, as illustrated schematically in Fig.~\ref{fig:pwave_pole_selfenergy}.
To illustrate the pole structure, we denote the one-loop self-energy of the intermediate baryon $B_n$ by $\Sigma_{B_n}(\slashed p)$. 
The corresponding Dyson-resummed propagator can be written schematically as\\
\begin{align}
S_{B_n}^{\rm full}(\slashed p)
&=
\frac{i}{\slashed p-m_0-\Sigma_{B_n}(\slashed p)}
\nonumber\\
&=
\frac{i}{
(\slashed p-m_{B_n})
\left[1-\Sigma'_{B_n}(m_{B_n})\right]
+\mathcal{O}\!\left((\slashed p-m_{B_n})^2\right)
}
\nonumber\\
&=
\frac{i}{\slashed p-m_{B_n}}
\frac{1}{
1-\Sigma'_{B_n}(m_{B_n})
+\mathcal{O}(\slashed p-m_{B_n})
}
\nonumber\\
&=
\frac{i}{\slashed p-m_{B_n}}
\left[
Z_{B_n}
+\mathcal{O}(\slashed p-m_{B_n})
\right]
\nonumber\\
&=
\frac{i Z_{B_n}}{\slashed p-m_{B_n}}
+
\frac{i\,\mathcal{O}(\slashed p-m_{B_n})}
{\slashed p-m_{B_n}} ,
\label{eq:resummed_pole_propagator}
\end{align}
where $m_0$ is the octet-baryon mass in the chiral limit, and $m_{B_n}$ is the physical mass of the intermediate baryon, determined by the pole condition
\begin{equation}
m_{B_n}-m_0-\Sigma_{B_n}(m_{B_n})=0 .
\end{equation}
The residue factor is given by
\begin{equation}
Z_{B_n}
=
\frac{1}{1-\Sigma'_{B_n}(m_{B_n})}.
\end{equation}

In the actual $P$-wave pole diagrams, the momentum flowing through the intermediate baryon propagator is fixed by the on-shell external baryon.
Therefore, after using the external equations of motion, the denominator $\slashed p-m_{B_n}$ reduces to
\begin{equation}
\slashed p-m_{B_n}
\longrightarrow
m_{B_i}-m_{B_n}
\qquad
\text{or}
\qquad
m_{B_f}-m_{B_n},
\end{equation}
depending on whether the intermediate propagator is attached to the initial or final baryon line. These mass differences are small $SU(3)$-breaking quantities. Hence, the first term in Eq.~\eqref{eq:resummed_pole_propagator} is enhanced by the small pole denominator, while the second term,
\begin{equation}
\frac{i\,\mathcal{O}(\slashed p-m_{B_n})}
{\slashed p-m_{B_n}},
\end{equation}
is finite in this limit and does not contain the pole enhancement. In the present prescription, we retain only the pole-enhanced part of the resummed propagator. 
The full resummed propagator can then be decomposed as 
\begin{equation}
\label{full_propa_decom}
    S_{B_n}^{\rm full}(\slashed p)=\frac{i}{\slashed{p}-m_{B_{n}}}+(Z_{B_{n}}-1)\frac{i}{\slashed{p}-m_{B_{n}}}.
\end{equation}
The first term on the right-hand side of Eq.~\eqref{full_propa_decom} represents the contribution that replaces the mass parameter in the pole-diagram propagator with the corresponding physical baryon mass. As discussed in the main text, this replacement is implemented in the $P$-wave pole diagrams for numerical stability. Consequently, the contribution from the first term has already been incorporated into the tree-level results in Eq.~\eqref{eq:tree_level_p_wave_amplitudes} and should not be included again explicitly. 
Thus, only the explicit contribution from the second term needs to be retained. This contribution is equal to the corresponding tree-level contribution multiplied by $(Z_{B_n}-1)$.

Explicitly, using Eq.~\eqref{full_propa_decom}, the relevant contributions
$\gamma^{\mathit{b+c},B_n}_{B_f B_i}$ from
Figs.~\ref{fig:pwave_pole_selfenergy}(b) and (c), where $B_n$ denotes the
intermediate baryon on the line dressed by the self-energy insertion, can be
written as
\begin{equation}
    \gamma^{\mathit{b+c},B_n}_{B_f B_i}
    =
    (Z_{B_n}-1)\alpha^{\mathit{a},B_n}_{B_f B_i}.
\end{equation}
Here, $\alpha^{\mathit{a},B_n}_{B_f B_i}$ denotes the contribution from
Fig.~\ref{fig:pwave_pole_selfenergy}(a) with the same intermediate baryon
$B_n$, while $B_i$ and $B_f$ denote the initial and final baryons,
respectively.

Finally, the total contribution corresponding to
Fig.~\ref{fig:loop_extra_pwave} is obtained by summing over all possible
intermediate baryons propagating along the baryon line dressed by the
self-energy insertion. This contribution is denoted by
$\gamma^{P}_{B_f B_i}$ in the total $P$-wave amplitude of
Eq.~\eqref{iM_P_structure}:
\begin{equation}
\begin{aligned}
&\gamma^{P}_{B_f B_i}
=
\sum_n
\left(
\gamma^{\mathit{o_1+p_1},B_n}_{B_f B_i}
+
\gamma^{\mathit{o_2+p_2},B_n}_{B_f B_i}
\right),
\\
&\gamma^{\mathit{o_1+p_1},B_n}_{B_f B_i}
=
\left(Z_{B_n}-1\right)
\alpha^{\mathit{b}_1,B_n}_{B_f B_i},
\\
&\gamma^{\mathit{o_2+p_2},B_n}_{B_f B_i}
=
\left(Z_{B_n}-1\right)
\alpha^{\mathit{b}_2,B_n}_{B_f B_i}.
\end{aligned}
\end{equation}
Results obtained using this prescription in the heavy-baryon framework can be
found in Refs.~\cite{Springer1995Revisited,BorasoyHolstein1999}.

\section{Unconsidered loop diagrams}
\label{app:unconsidered_p2_loops}
\begin{figure}[htbp]
\centering

\includegraphics[
  width=0.6\textwidth,
  height=0.82\textheight,
  keepaspectratio
]{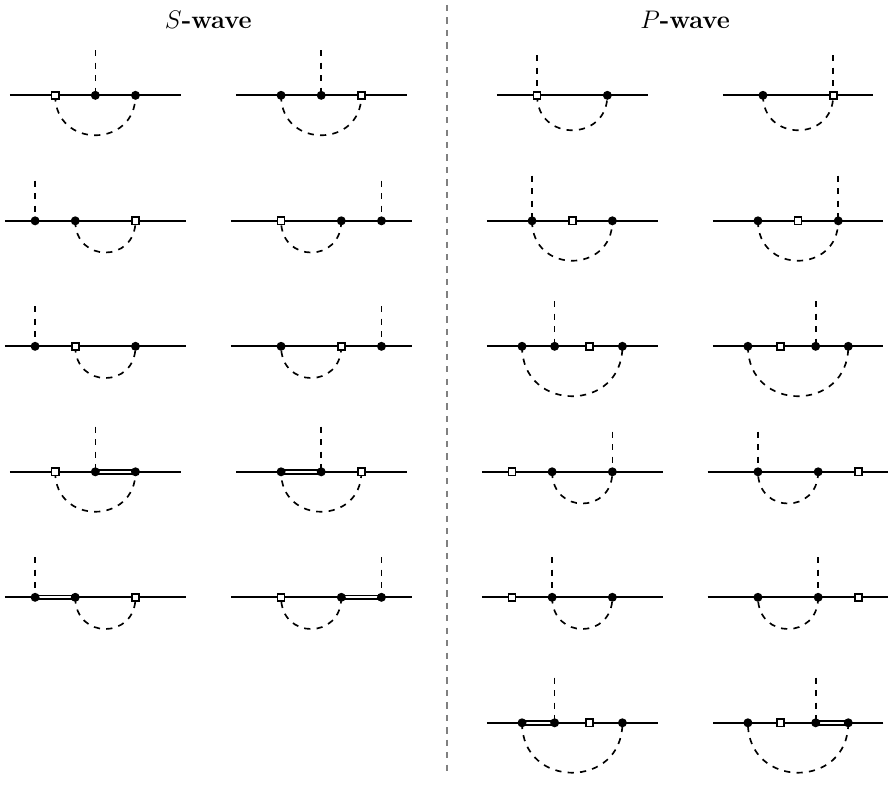}

\caption{Feynman diagrams for $\Delta S=1$ hyperon decay that are not considered in this work. Left panel: $S$-wave amplitudes. Right panel: $P$-wave amplitudes.}
\label{fig:loop_unconsidered_combined}
\end{figure}

This section displays the NLO loop diagrams that are not considered in the
present work. The corresponding $S$-wave and $P$-wave topologies are shown in
the left and right panels of Fig.~\ref{fig:loop_unconsidered_combined},
respectively.

\clearpage
\twocolumngrid
\bibliographystyle{apsrev4-2}
\bibliography{references}
\end{document}